\documentclass[11pt,a4paper]{article}
\usepackage[latin1]{inputenc}
\usepackage{amsmath}
\usepackage{amsthm}
\usepackage{amsfonts}
\usepackage{amsfonts,amsthm,latexsym,amsmath,amssymb,amscd,epsfig,psfrag,enumerate}
\usepackage{graphics,graphicx, bezier, float, color, hyperref}
\usepackage{amssymb,url}
\usepackage{multienum}%
\usepackage[table]{xcolor}
\usepackage{multicol,multirow}
\usepackage{graphicx}
\usepackage{fancyvrb}
\usepackage{parskip}
\usepackage[toc,page]{appendix}

\usepackage{hyperref}  
\usepackage{float}

\usepackage{subcaption} 
\usepackage{pdfpages}
\usepackage[top=2.7cm, bottom=2.7cm, left=1.5cm, right=1.5cm]{geometry}
\usepackage{xcolor}
\usepackage{lipsum}
\usepackage{blkarray}
\newtheorem{conjecture}{Conjecture}

\usepackage[none]{hyphenat}[section]

\numberwithin{equation}{section}
\numberwithin{table}{section}
\numberwithin{figure}{section}
\usepackage{multirow}
\color{black}  
\usepackage{mathptmx}  

\title{\textbf{	Hedging Options on Asset Portfolios against Just One Underlying Asset in the Presence of Transaction Costs}} 
\author{Erina Nanyonga$^{\text{*}} $, Matt Davison$^{\text{**}} $\\
	$^{\text{*}}$\textit{Department of Mathematics  Makerere University},\\$^{ \text{**}}$	\textit{Department of Statistical and Actuarial Sciences
		Western University Canada}\\
	Corresponding author's email: \textit{erinananyonga9@gmail.com}}
\date{}
\begin{document}
	
	\maketitle
	\begin{abstract}
		\noindent
		\textbf{Abstract}
		
		\noindent
		Options are contingent claims regarding the value of underlying assets.  The Black-Scholes formula provides a road map for pricing these options in a risk-neutral setting,  justified by a delta hedging argument in which countervailing positions of appropriate size are taken in the underlying asset.   However, what if an underlying asset is expensive to trade?  It might be better to hedge with a different, but related asset that is cheaper to trade. This study considers this question in a setting in which the option written on a  portfolio containing $\alpha$ shares of one asset $S_{t_1}$  and $(1-\alpha)$ shares of another security $S_{t_2}$ correlated with $S_{t_1}$.  We suppose that the asset is hedged against only one of $S_{t_1}$ or $S_{t_2}.$  In the case of $\alpha=0~\text{or}~1$  we can consider this model to cover the case where an option on one asset is hedged against either the ``right" (underlying) asset or the``wrong" (related, different) asset.  We hedge our portfolio on simulated data using varying trading intervals, correlation coefficients, $\rho$ and  transaction costs. We calculated the risk-adjusted values ($RAV$) as the risk and return measures to make meaningful decisions on when to trade $S_{t_1}$ or $S_{t_2}.$  From the conclusions made based  on $RAV,$ the  size of the market price of risk and that  of transaction costs on both  assets are key to making a decision while hedging. From our results, trading the wrong asset can be opted for when $\rho$ is very high for  reasonably small transaction costs for either of the assets. 
		
		\vspace{1em}
		\noindent\text{Keywords:} Hedging; transaction costs; correlation; equity options; risk-adjusted value; unhedged value; rebalancing intervals; Monte Carlo simulations, Leland's number
	\end{abstract}	
	
	\setlength{\parindent}{20pt}  
	\setlength{\parskip}{0pt}     
	
	\vspace{0.2cm}
	\section{Introduction}

	Derivatives are securities whose values are derived from the unknown future value of another ``underlying" security.  This implies a mathematical relationship between the value of a derivative and the value of the other security; this relationship is exploited to create hedging strategies that ideally eliminate, but at least reduce, price risk.  The argument leading to the famous Black-Scholes equation relies on a continuous hedging strategy which in the simplified model of reality used,  allows the creation of a hedging strategy that completely eliminates risk.  In reality,  market imperfections such as transaction costs, multiple risk drivers, and periodic inabilities to trade mean that practical hedging does not eliminate risk completely, but  allows it to be managed.
	
	Hedging is costly because trading in the underlying asset incurs costs both in the form of brokerage fees and  slippage, which measures the degree to which buying a security pushes up its value and selling it pushes down its value.   Thus,   investors must trade the frequency of rebalancing against the cost of rebalancing.  If trading in the underlying asset is sufficiently costly,  it might even be better not to hedge.  This alternative, termed as holding a naked position,  incurs no trading costs but also fails to eliminate any risk. 
	
	The selection of an appropriate investment strategy depends on investors' choice  to achieve  their target objectives that will cater to any involved transaction costs during investment \cite{steuer2003multiple}. 
	Disregarding transaction costs would lead to an inefficient portfolio  \cite{arnott1990measurement}, whereas including transaction
	costs would assist decision makers in  understanding the nature of an efficient portfolio \cite{sadjadi2004dynamic}. 
	Transaction costs may play a significant role in diminishing arbitrage
	profits by hindering an investor from purchasing an under priced asset with the intention of selling it a few weeks or months later, and
	recognizing a profit net of transaction costs \cite{constantinides1986capital}. The presence of transaction costs leads to a reduction in the derivative values, and this reduction depends on the interval at which the derivative is rebalanced. The transaction costs are composed of a fixed part and a proportional part, which are affected
	by several factors. The main factor affecting the size of transaction costs is slippage, which is proportional to the amount traded but often quantified by the  bid-ask spread  \cite{gemmill1993options}.

	The practical research question addressed in this study is
	against which asset blend, how often, and when one should hedge a portfolio. While frequent   option rebalancing  reduces hedging errors and leads to high transaction costs, less frequent rebalancing leads to large hedging errors.
	Rebalancing arises from consistent portfolio monitoring and involves a set of trading rules designed to adjust the portfolio to align it with its objectives and constraints. When transaction costs are involved, rebalancing must consider the deviation from the target and the expenses required to minimize that gap \cite{cesari2012trading}. 
	Investors  consider the expected return and risk during the  rebalancing process. 
	In the presence of transaction costs, investors reduce  the frequency of trading because the more one trade, the more  money is lost. 
	The need for
	rebalancing depends on asset characteristics including \cite{cesari2012trading}; first, volatility, expected return,
	and time in a direct way, because they imply a high probability of significant deviations of one asset relative to the
	other.
	Second, correlations of assets in an inverse way, as high correlations among assets, imply strong comovements, which reduce the need for rebalancing.

	In this study, we  carried out delta hedging of  equity options on geometric Brownian motion (gBm) assets,  including and excluding  transaction costs.  We assumed proportional transaction costs and practice discrete-time hedging.  
	In the absence of transaction costs, the Black-Scholes delta hedging strategy performs sufficiently well compared to hedging strategies given by more complex models. The famous Black-Scholes hedging strategy performs perfectly for equity options written on a gBm asset that can be traded  continuously and without transaction costs;  it continues to perform very well, even if trading is only done at a positive rebalancing interval. %
	
	There have been modifications to the Black-Scholes theory of hedging, which involves transaction costs. In 1985, Leland published a paper on option hedging with transaction costs,  which was an extension of the Black-Scholes model \cite{leland1985option}. In 1992, Boyle and Vorst considered option hedging with transaction costs as an extension of the Cox, Ross, and Rubinstein model \cite{boyle1992option}.
	Although delta is well analyzed, it
	does not eliminate all risks. According to  \cite{neuberger1994log}, a trader may be subjected to losses if volatility spikes suddenly even with a properly delta hedged portfolio. 
	
	Hedging in the Black-Scholes strategy must be done continuously while holding the delta (shares) of the asset. 
	The gBm model,  utilized in this study  to simulate asset prices, is characterized by infinite variation; thus, for any non-zero transaction costs, the
	continuous replication policy mandated by the Black-Scholes model incurs
	an infinite amount of transaction costs over any trading interval, regardless of
	how small it might be \cite{zakamouline2006optimal}. 
	Other strategies include Leland's strategy,
	delta tolerance strategy, asset tolerance strategy, and utility maximization-based strategy. 	\cite{clewlow1997optimal} studied the problem of delta hedging portfolios of options under transaction costs by maximizing expected utility. Their results 
	showed that the optimal control approach is substantially more effective than discrete rebalancing strategies such as the standard implementation of
	Black-Scholes or Leland's method. 
	\cite{marco1994dynamic} reported that the type of hedging strategy to be used depends on the value of the Leland number A, ~$\text{A}=\sqrt{\frac{2}{\pi}\frac{k}{\sigma \sqrt{\delta t}}},$ where $k$ is the 
	transaction cost, $\sigma$ is the volatility of the underlying asset, and $\delta$t is the time lag between transactions. If A $<$ 1, it is possible to apply modified Black-Scholes delta hedging strategies, but not otherwise. 
	
	\cite{mohamed1994simulations}
	evaluates several rehedging strategies using Monte Carlo simulations. The Black-Scholes environment assumes frictionless markets in which one can replicate the option payoff exactly by continuous rehedging.  In contrast to some of our simulations, it also assumed a  risk free scenario.
	However, when transaction costs are involved, frequent rehedging results in the accumulation of transaction costs \cite{mohamed1994simulations},  and he argues that infrequent hedging results in replication errors.  In this study, we conducted Monte Carlo simulations on  simulated  values in the presence and absence of varying transaction costs at different rebalancing intervals.

	\cite{martellini2002competing} note that for all hedging strategies, the higher the volatility, the higher the transaction costs incurred. This is true because the dynamic replication of the option payoff constitutes   more transactions when  hedging volatility  increases. In this study, we calculate the risk-adjusted value as a risk and return measure.
	The paper proceeds as follows. Section \ref{sec:2} introduces the hedging procedure that we considered to rebalance our portfolio. It presents a procedure for hedging one of the assets in our portfolio. Section \ref{sec:3}  presents the numerical results of our hedging strategy and describes the managerial insights of the hedging strategy. Finally, Section \ref{sec:4} concludes the study.

	\section{Hedging Procedure}\label{sec:2}
	Delta hedging   is  a method that is expensive in terms of transaction costs, because it requires frequent
	rebalancing. Every move in price leads to a different delta, and consequently, the portfolio must be rebalanced by selling or buying additional shares to stay delta-neutral. Many authors investigate delta hedging not only from the Black-Scholes model point of view but also by using
	other stochastic models.
	According to \cite{zakamouline2006optimal}, 
	the minimum or maximum number of rebalancing intervals is not yet clear  from previous  studies on delta hedging. 
	
	The partial differential equation (pde) in the absence of transaction costs that satisfy the number of shares, $\Delta_t$ held in an asset is given by
	\[
	\begin{cases} 
		-rC_t + \frac{\partial C_t}{\partial t}+rS_t\frac{\partial C_t}{\partial S_t}+\frac{1}{2}\sigma^2 S_t^2\frac{\partial^2 C_t}{\partial^2 S_t}& =0 \\ 
		\Delta_t & = \frac{\partial C_t}{\partial S_t} 
	\end{cases}
	\] for  asset price $S_t$ and  option price $C_t$ at time $t$. The risk-free interest rate is denoted by $r$ and the volatility of the asset is given by $\sigma.$
	The final payoff at  expiry time $T$ is   $C_T=\text{Max}(S_T-K,0)$, where $S_T$ is the final portfolio price and $K$ is the strike price.
	With the inclusion of transaction costs, the pde approach above is not available because of  dependence of the change in shares of the underlying asset on the change in time. However, a similar pde with an adjusted volatility can be employed in the presence of transaction costs \cite{wilmott1994hedging}.  
	The presence of transaction costs invalidates the assumptions of the Black-Scholes 
	model \cite{kocinski2014transaction}.
	However, although the Black Scholes model has incorrect assumptions, it still works
	remarkably well \cite{black1989use}. 
	
	In this study, we delta hedged  a  portfolio of two assets and  traded only one of them in the entire portfolio.  A portfolio, $S_t$ of two assets with $\alpha$ shares in the first asset, $S_{t_1}$ and $(1 -\alpha)$  shares in the second asset, $S_{t_2}$ was  considered, as described by Equation \eqref{eqn1}:
	\begin{equation}\label{eqn1}
		S_t=\alpha S_{t_1} +(1-\alpha)S_{t_2},
	\end{equation}
	While trading only one of the assets, one asset was at times not perfectly correlated to the other but cheaper to trade; this allowed us to smoothly perturb our strategy from trading in the correct, but more expensive to rebalance asset, to trading only in the incorrect, but cheaper to rebalance asset.  In Equation \eqref{eqn1}, the portfolio is strictly on  $S_{t_1}$  when $\alpha =0$ and on  $S_{t_2}$ when $\alpha=1$. Other $\alpha$ values besides 1 and 0 make the portfolio blended, with each asset contributing a certain percentage.

	\subsection{Trading just one of the assets in our portfolio}
	The portfolio  considered in the rest of this paper is  described by Equation \eqref{eqn1}. We considered a scenario in which  it is  possible to trade either $S_{t_1}$ or  $S_{t_2}$ at expiry, but not both, in the portfolio. We rebalanced one of the assets in the portfolio at each time and traded one  of them at expiry in the presence and absence of transaction costs on the simulated data. 
	
	Both $S_{t_1}$ and $S_{t_2}$ were considered to follow gBm with correlated Brownian innovations. Equation \eqref{eqn1} is a sum of the two assets hence, not a gBm.
	For the first asset,  we have 
	\begin{equation}\label{eqn2}
		dS_{t_1}=\mu_1 S_{t_1}dt + \sigma_1 S_{t_1}dW_1(t),
	\end{equation} 
	while for the second asset we have 
	\begin{equation}\label{eqn3}
		dS_{t_2}=\mu_2 S_{t_2}dt + \sigma_2 S_{t_2}dW_2(t),
	\end{equation}
	where $\sigma_1$ and $\sigma_2$ are the annualized volatilities of assets $S_{t_1}$ and $S_{t_2}$ respectively 
	$\mu_1$ and $\mu_2$ are the drift terms of $S_{t_1}$ and $S_{t_2}$  respectively; and  $W_1$ and $W_2$ are the Wiener processes. Equations \eqref{eqn2} and \eqref{eqn3}  correlate with  $\mathbb{E}[dW_1(t)dW_2(t)]=\rho dt$, where $\rho$ represents the correlation coefficient of the two assets in the portfolio. The discrete time approximations of the gBm equations, given $T$ as the expiry time are;
	
	\begin{equation}\label{eqn4}
		S_{t_1}(t) = S_{t_1} (t-1)\exp\left( \left(\mu_1 - \frac{1}{2} \sigma_1^2 \right) d t + \sigma_1 \sqrt{d t} Z_{1_t} \right),
	\end{equation}
	
	\begin{equation}\label{eqn5}
		S_{t_2}(t) = S_{t_2} (t-1)\exp\left( \left(\mu_2- \frac{1}{2} \sigma_2^2 \right) d t + \sigma_2 \sqrt{d t} Z_{2_t} \right),
	\end{equation}
	Where $Z_{1}$ and  $Z_{2}$ are two correlated Wiener processes with 
	$Z_{1} \sim \mathcal{N}(0, 1) $ is a random variable drawn from a standard normal distribution  representing the Brownian motion,  and in our simulations, \begin{equation}\label{eqn6}Z_{2}= \rho * Z_{1} + \sqrt{1 - \rho^2}~W, 
	\end{equation}
	where $W$ is another random variable,  $W \sim \mathcal{N}(0, 1).$ 
	The initial portfolio price, $S_{t_0}$ is given by 
	\begin{equation}\label{eqn7}
		S_{t_0}=\alpha S_{t_{1_0}}+(1-\alpha)S_{t_{2_0}},
	\end{equation} 
	where  $S_{t_{1_0}}$ and $S_{t_{2_0}}$ are the initial prices of $S_{t_1}$ and $S_{t_2}$ respectively.
	We considered
	$ S_{t}= S_{t_1}, ~~   S_{t}=St_{2},
	$
	to be the price of the portfolio at time $t$ while trading only $S_{t_1}$ and $S_{t_2}$ respectively.
	The entire portfolio has a variance given by
	$$ \sigma^2 = \alpha^2\sigma_1^2 + (1-\alpha)^2\sigma_2^2 + 2\alpha(1-\alpha) \rho \sigma_1 \sigma_2 ,$$
	where $\sigma_1$ and $\sigma_2$ are the volatilities of the individual assets, and $\rho$ is the correlation coefficient between the two assets.
	
	While trading only  $S_{t_1}$ in our portfolio in the presence of transaction costs of the first asset, $k_{1}$ is  the  initial holding   $\Delta_{1_0}=\frac{\partial C_{1_0}}{\partial S_{t_{1_0}}},$ and 
	represents the initial shares for an investor while trading only the first asset, where $C_{1_0}$ represents the initial option price while trading only asset 1.  $S_{t_{1_0}}$ is the initial stock price of $S_{t_1}$. At each rebalancing time $t,$ a new share value $\Delta_{1_t}$  is computed using the stock price of the first asset at that time.   $\Delta_{1_t}$ is also given by; $\Delta_{1_t}=\frac{\partial C_{t_1}}{\partial S_{t_1}}=\phi(d_{1_1})$,  where $\phi$ is the cumulative normal distribution and $C_{t_1}$ is the option price when only asset 1 is  traded  at time $t$.  The $d_{1_1}$ for updating $\Delta_{1_t}$  is given by
	\begin{eqnarray}\label{eqn19}
		d_{1_1}&=& \frac{\ln\left(\frac{S_{t_1}}{K}\right) + \left(r + \frac{\sigma_1^2}{2}\right)\tau}{\sigma_1 \sqrt{\tau}},
	\end{eqnarray} where $\tau$ is the time remaining  to expiry (initial $\tau$~$=1=T$). In the case of trading only the second asset, the initial holding $\Delta_{2_0}=\frac{\partial C_{2_0}}{\partial S_{t_{2_0}}}$, where $C_{2_0}$ is the initial option price when asset 2 is the only one traded in the portfolio and $S_{t_{2_0}}$ is the initial stock price of the second asset. The new shares while trading only $S_{t_2}$ are given by $\phi(d_{1_2})$ with 
	
	\begin{eqnarray}\label{eqn20}
		d_{1_2}&=& \frac{\ln\left(\frac{S_{t_2}}{K}\right) + \left(r + \frac{\sigma_2^2}{2}\right)\tau}{\sigma_2 \sqrt{\tau}},
	\end{eqnarray} 
	
	The Black-Scholes price for the portfolio using the portfolio variance and the initial portfolio price, $S_{t_0}$ was computed  to compare the simulated price with the theoretical  Black-Scholes  price described by  
	\begin{equation}\label{eqn10}
		C_{BS} = St_0 \Phi(d_1) - Ke^{-rT} \Phi(d_2),
	\end{equation}
	\begin{eqnarray*}\text{where}~~
		d_1 &=& \frac{\ln\left(\frac{St_0}{K}\right) + \left(r + \frac{\sigma^2}{2}\right)T}{\sigma \sqrt{T}}~~\text{and}~~
		d_2 =d_1 - \sigma \sqrt{T}.
	\end{eqnarray*} 
	
	Let $PV_1$ and $PV_2$ represent the portfolio value when trading only $S_{t_1}$ and $S_{t_2}$ respectively, in the portfolio.
	When only the first  asset is traded,  the initial portfolio value in the presence of transaction costs of the first asset, $k_1$ is 
	\begin{eqnarray}\label{eqn11}
		\text{Initial} ~PV_1&=&S_{t_{1_0}}\Delta_{1_0} (1-k_1),\\\nonumber
		&=&C_{1_0} - k_1~\Delta_{1_0}  S_{t_{1_0}}.
	\end{eqnarray}
	For the case of trading only the second asset with transaction costs of the second asset, $k_2;$ 
	\begin{eqnarray}\label{eqn12}
		\text{Initial} ~PV_2&=&S_{t_{2_0}}\Delta_{2_0} (1-k_2)\\\nonumber
		&=&C_{2_0} - k_2~\Delta_{2_0}  S_{t_{2_0}}.
	\end{eqnarray} The portfolio value at each time point was adjusted using the delta value. For asset 1, the current portfolio value at time $t$ is
	\begin{equation}\label{eqn13}\text{Current }~PV_{1}= (1 + r dt)  (\text{previous~~} PV_{1}) + (\Delta _{1_t}- \Delta _{1_{(t-1)}}S_{t} \Bigg(1 - \text{sign}\left[\Delta _{1_t}- \Delta _{1_{(t-1)}}\right] k_1\Bigg),
	\end{equation}
	
	where  $\Delta _{1_t}$ is  an investor's holding at time $t$, $S_t$ is the current portfolio price, $    S_{t}= S_{t_{1}}$, $r$ is the risk-free interest rate, $k_1$ is the transaction cost of the first asset.  At expiry,  the final price of the portfolio is given by
	\begin{equation}\label{eqn14}
		S_T = \alpha S_{T_1}+(1-\alpha) S_{T_2},
	\end{equation}
	where $S_{T_1}$ and $S_{T_2}$ are the final simulated prices of the first and second asset respectively.
	The final portfolio value while trading only the first asset is given by Equation \eqref{eqn15}
	
	\begin{eqnarray}\nonumber
		\text{Final }~PV_{1}&=&  (1 + r dt)  ~\text{previous}~ PV_{1} + (\Delta _{1_T}- \Delta _{1_{(t-1)}})S_{T_1}\Bigg(1 - \text{sign}\left[\Delta _{1_T}- \Delta _{1_{(t-1)}}\right] k_1\Bigg) \\\label{eqn15}&&\hspace{6cm} - \Delta_{1_T} S_{T_1} + \text{Max}(S_T - K, 0),
	\end{eqnarray}
	
	where $K$ is the strike price, $S_{T_1} $ is the final price of asset 1 at expiry, $\Delta _{1_T}$ is the final delta value of the final price of the first asset (asset 1). 
	In the case of trading only the second asset;
	\begin{equation}\text{Current }~PV_{2}= (1 + r dt)  ~\text{previous}~ PV_{2} + (\Delta _{2_t}- \Delta _{2_{(t-1)}}) S_{t} \Bigg(1 - \text{sign}\left[\Delta _{2_t}- \Delta _{2_{(t-1)}}\right] k_2\Bigg),
	\end{equation}
	
	where  $\Delta _{2_t}$ is the share of the second asset at time $t,$  $  S_{t}$ is the current portfolio price  $=St_{2}$, $r$ is the risk-free interest rate.  
	The final portfolio value while trading only asset 2 is  given by Equation \eqref{eqn17}:
	\begin{eqnarray}\nonumber
		\text{Final }~PV_{2}&= & (1 + r dt)  ~\text{previous}~ PV_{2} + (\Delta _{2_T}- \Delta _{2_{(t-1)}}) S_{T_2}\Bigg(1 - \text{sign}\left[\Delta _{T_2}- \Delta _{2_{(t-1)}}\right] k_2\Bigg)\\\label{eqn17}&&\hspace{6cm} - \Delta_{2_T} S_{T_2} + \text{Max}(S_T - K, 0),
	\end{eqnarray}
	
	where $K$ is the strike price, $S_{T_2} $ is the final  price of the second asset (asset 2), $\Delta _{2_T}$ is the final delta value for the final price of the second asset,  and the other parameters are as previously described. 
	The simulated portfolio price is obtained by discounting Equations \eqref{eqn15} and \eqref{eqn17} at expiry. This price represents what is available to the hedger after  hedging (the return from hedging)
	\begin{equation}\label{eqn18} \text{simulated portfolio price }= e^{-r T} ( \text{Final}~PV),
	\end{equation}where $PV$ represents either $PV_{1}$ when only asset 1 is traded or $PV_{2}$ when asset 2 only is traded.
	The unhedged value was  computed  to  compare how close the simulated portfolio prices were to the theoretical expected portfolio value at expiry if  there was no hedging at all.  
	\text{The unhedged value is given as;}
	\begin{equation}\label{eqn19}
		C_{u.v}(S,t)=e^{-r(T-t)}\left(S_t( e^{\mu (T-t)} \phi(d_1))-K(\phi(d_2)\right),
	\end{equation}
	
	where $\phi (d_1)$ and $\phi (d_2)$ are the cumulative normal distributions of $d_1$ and $d_2$ respectively.
	This expression is the Black-Scholes price when $\mu = r$.  The value of $\mu$ is the portfolio drift given by
	\begin{equation}\label{eqn20}
		\mu = \alpha \mu_1+(1-\alpha)\mu_2,
	\end{equation}          
	where $\mu_1$ and $\mu_2$ are the drifts of $S_{t_1}$ and $S_{t_2}$ respectively.
	We also computed the measure value at expiry as
	\begin{equation}\label{eqn21}
		\text{Measure }= \text{exp}~(-r T)~  C_T,
	\end{equation} 
	where    $C_T = \text{Max}(S_T - K, 0)$ denotes the realized option price.  
	The measure value was the simulated unhedged portfolio value. The averages of the simulated measure values were considered as the discounted $P$ measure value, which was considered to be the simulated unhedged value of the portfolio.

	\subsection{Risk-adjusted value ($RAV$)}
	
	When designing a portfolio, it is essential to consider a client's risk tolerance and return expectations to maximize the expected returns for a given level of risk. Modern portfolio theory (MPT),  invented by Markowitz in 1952 \cite{markowitz1952modern} selects stocks for a portfolio that accounts for the trade-off between their risk premium (return above the risk-free rate) and their volatility, and for both correlations with other assets.
	However, despite its widespread application and recognition, MPT has notable shortcomings due to its unrealistic assumptions and its failure to account for macroeconomic factors and  potential changes in company performance \cite{yu2023comprehensive}.

	$RAV$ evaluates the return on an investment relative to the  associated risk. 
	Various methods exist for measuring risk-adjusted returns, including ratios and models that account for different types of risks.
	According to \cite{crouhy1999measuring},
	performance metrics such as return on capital have been widely acknowledged for neglecting  the risk associated with the underlying business and the value of future cash flows.
	Investors use risk-adjusted returns to compare potential investments and make informed decisions by weighing the potential returns against inherent risks.
	
	The mean and variance as measures of risk can be summarized in the Sharpe ratio, which  measures excess return per unit of risk where risk is measured by
	the standard deviation of the excess returns 
	\cite{sharpe1994sharpe}. 
	We  consider the $RAV$ in the case of trading only one asset  as the risk and return measure.
	\begin{equation}\label{eqn22}RAV=\text{average simulated value}-  \lambda(\text{Standard deviation of the simulated value}),\end{equation}
	$\text{where}~\text{the market price of risk,} ~\lambda=\frac{\left(\mu_r - r \right)}{\sigma_r},$
	where $\mu_r$ is the expected return, $r$ is the risk-free interest rate, and $\sigma_r$ is the standard deviation of $\mu_r.$ The maximum value of  $RAV,$ is the best decision to opt for.  $RAV$ represents the return above  the risk-free rate that an investor  expects for every unit of risk. 
	
	\cite{marco1994dynamic} concluded that if  Leland's number, A $<$ 1 it is possible to apply modified Black-Scholes delta hedging strategies, but not otherwise.  We computed the  Leland number A, for the modified Black-Scholes strategy that we applied in presence of transaction costs used as;  ~$\text{A}=\sqrt{\frac{2}{\pi}\frac{k}{\sigma \sqrt{\delta t}}},$ where $k$ is the  transaction cost, $\sigma$ is the volatility of the underlying asset, and $\delta t$ is the time-lag between transactions and from our results, not hedging at all is always better than hedging when $A>1.$
	
	\section{ Numerical Results}\label{sec:3}
	
	We conducted Monte Carlo simulations,  where the average of all the simulated portfolio prices for the runs used was considered as the simulated portfolio price for each $\alpha$ value.  We calculated the average of the following simulated values: the simulated portfolio price, final portfolio price, measure value, Black-Scholes price, unhedged value, and their standard deviations for each $\alpha$ value in each simulation. We considered the averages of the runs  as the values needed for each value of $\alpha$.   The following are the definitions of the  parameters for our simulations: 
	
	$S0_1$ is the initial price of $S_{t_1}$, ~
	$S0_2$  is the initial price of $S_{t_2}$,~
	%
	$\rho$ is the correlation coefficient between the two assets;~
	$K$  is the strike price of the option;~
	$r$ is the risk-free interest rate;~
	%
	$k_1$ is the transaction cost of $S_{t_1},$~
	$k_2$ is the transaction cost of $S_{t_2},$~
	$\sigma_1$ is the volatility of $S_{t_1},$~
	$\sigma_2 $ is the volatility of $S_{t_2},$~
	$T = 1 $  is the time  (year) to expiration;~
	$\mu_1$ is the drift rate of $S_{t_1},$~
	$\mu_2$ is the drift rate of $S_{t_2},$~
	num\_steps  is the number of rebalancing intervals (number of steps); and 
	~num\_simulations are the number of simulations.

	In Subsection \ref{subsec:3.1}, we have some of our results in tables where we let $\alpha$ represent the proportion of assets in Equation \eqref{eqn1},  BS price  represents the  Black-Scholes price,  avg sim price  represents the (average) simulated portfolio price,  std\_sim price  represents the average of the  standard deviation of the  simulated portfolio price in the simulation,  avg S\_T  represents the average final portfolio price at expiry,  std\_S\_T  represent the average of the  standard deviation of the final portfolio price at expiry,  dis P\_mea represents the discounted $P$ measure value (the average of the measure values), std\_dis\_P  represents the average of the standard deviation of the measure values, and Avg\_unhedged represents the average of the unhedged values for each value of $\alpha$. 
	\subsection{Simulations for  hedging one of the assets in Equation \eqref{eqn1} }\label{subsec:3.1}
	
	\subsubsection{Case I: Trading only $S_{t_1}$;}\label{subsubsec:3.1.1}
	
\begin{figure}[htbp]
		\centering
		\subfloat[$k_1=0.0$.]{%
			\resizebox*{7cm}{!}{\includegraphics{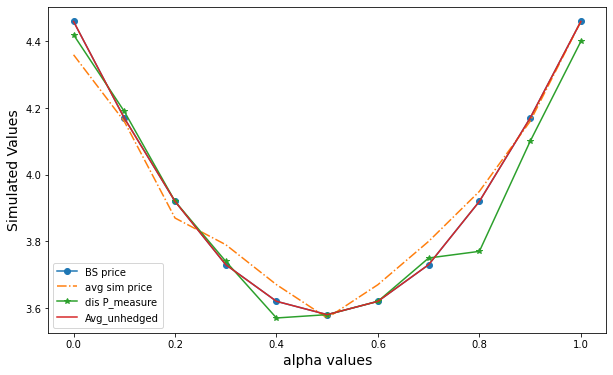}}}\hspace{5pt}
		\hspace{5pt}
		\subfloat[$k_1=0.005$.]{%
			\resizebox*{7cm}{!}{\includegraphics{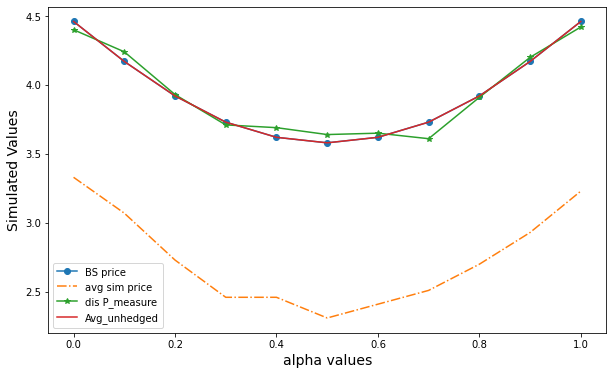}}}\hspace{5pt}
		
		\caption{  Trading only the first asset in the portfolio in absence and presence of transaction costs}  
		\label{fig1}
	\end{figure}

	\underline{Simulation I}.~~
	The following parameters were considered;
	$S0_1 = 50 ,~S0_2 = 50,~\rho = 0.2,~K = 50,~r = 0.02,~ T = 1 ,~\sigma_1 = 0.2, ~\sigma_2 = 0.2,~\mu_1 = 0.02,~\mu_2 = 0.02,~$ num\_steps $=252$, ~
	num\_simulations $= 10000,~k_1=0.005$, $\alpha$= $0 - 11.$ Leland's number  in this simulation:  A$<1$. 
	The experiment was conducted in the presence and absence of transaction costs on the traded asset to determine the effects of transaction costs during the hedging of only one of the assets in the whole portfolio. The simulation results are presented in Table 
	\ref{tab1}  and Figure \ref{fig1}.
	
	\begin{table}[h]\tiny
		\caption{Without and with  transaction costs of  $S_{t_1}$.}
		{			\resizebox{\columnwidth}{!}{
				\begin{tabular}{ccccccccccccccccc}
					\hline
					\multicolumn{3}{c}{}                                                                                                        & \multirow{2}{*}{} & \multicolumn{6}{c}{\text{Without Transaction costs}}                                                                                                                                                                                                                                                                       & \multirow{2}{*}{} & \multicolumn{6}{c}{\text{With Transaction costs}}                                                                                                                                                                                                                                                                          \\  \cline{5-10} \cline{12-17} 
					\text{$\alpha$} & \begin{tabular}[c]{@{}c@{}}\text{BS}\\  \text{price}\end{tabular} & \begin{tabular}[c]{@{}c@{}}\text{Avg \_}\\   \text{ unhedged}\end{tabular} &                   & \begin{tabular}[c]{@{}c@{}}\text{avg sim} \\ \text{price}\end{tabular} & \begin{tabular}[c]{@{}c@{}}\text{std \_}\\ \text{sim price}\end{tabular} & \begin{tabular}[c]{@{}c@{}}\text{avg}\\  \text{S\_T}\end{tabular} & \text{std \_S\_T} & \begin{tabular}[c]{@{}c@{}}\text{dis P\_}\\ \text{mea}\\  \text{sure}\end{tabular} & \begin{tabular}[c]{@{}c@{}}\text{std \_}\\ \text{dis \_P}\end{tabular} &                   & \begin{tabular}[c]{@{}c@{}}\text{avg sim}\\  \text{price}\end{tabular} & \begin{tabular}[c]{@{}c@{}}\text{std \_}\\ \text{sim price}\end{tabular} & \begin{tabular}[c]{@{}c@{}}\text{avg}\\  \text{S\_T}\end{tabular} & \text{std \_S\_T }& \begin{tabular}[c]{@{}c@{}}\text{dis P\_}\\ \text{mea}\\  \text{sure}\end{tabular} & \begin{tabular}[c]{@{}c@{}}\text{std \_}\\ \text{dis \_P}\end{tabular} \\ \hline
					0        & 4.46                                                & 4.46                                                       &                   & 4.36                                                     & 8.86                                                       & 50.96                                               & 10.27      & 4.42                                                          & 6.86                                                     &                   & 3.33                                                     & 8.73                                                       & 50.91                                               & 10.28      & 4.4                                                           & 6.83                                                     \\
					0.1      & 4.17                                                & 4.17                                                       &                   & 4.16                                                     & 8.05                                                       & 51.05                                               & 9.52       & 4.19                                                          & 6.38                                                     &                   & 3.07                                                     & 7.95                                                       & 51.12                                               & 9.53       & 4.24                                                          & 6.38                                                     \\
					0.2      & 3.92                                                & 3.92                                                       &                   & 3.87                                                     & 7.31                                                       & 51.05                                               & 8.86       & 3.92                                                          & 5.94                                                     &                   & 2.73                                                     & 7.29                                                       & 50.98                                               & 8.98       & 3.93                                                          & 6.04                                                     \\
					0.3      & 3.73                                                & 3.73                                                       &                   & 3.79                                                     & 6.4                                                        & 51.05                                               & 8.37       & 3.74                                                          & 5.58                                                     &                   & 2.46                                                     & 6.28                                                       & 51.05                                               & 8.29       & 3.71                                                          & 5.55                                                     \\
					0.4      & 3.62                                                & 3.62                                                       &                   & 3.67                                                     & 5.44                                                       & 50.95                                               & 8.01       & 3.57                                                          & 5.26                                                     &                   & 2.46                                                     & 5.43                                                       & 51.05                                               & 8.16       & 3.69                                                          & 5.42                                                     \\
					0.5      & 3.58                                                & 3.58                                                       &                   & 3.57                                                     & 4.49                                                       & 51.04                                               & 7.91       & 3.58                                                          & 5.22                                                     &                   & 2.31                                                     & 4.52                                                       & 51.13                                               & 7.97       & 3.64                                                          & 5.37                                                     \\
					0.6      & 3.62                                                & 3.62                                                       &                   & 3.67                                                     & 3.61                                                       & 50.97                                               & 8.07       & 3.62                                                          & 5.32                                                     &                   & 2.41                                                     & 3.58                                                       & 51.01                                               & 8.08       & 3.65                                                          & 5.33                                                     \\
					0.7      & 3.73                                                & 3.73                                                       &                   & 3.8                                                      & 2.71                                                       & 51.05                                               & 8.32       & 3.75                                                          & 5.56                                                     &                   & 2.51                                                     & 2.67                                                       & 50.87                                               & 8.29       & 3.61                                                          & 5.5                                                      \\
					0.8      & 3.92                                                & 3.92                                                       &                   & 3.95                                                     & 1.84                                                       & 50.78                                               & 8.79       & 3.77                                                          & 5.84                                                     &                   & 2.7                                                      & 1.85                                                       & 51.03                                               & 8.79       & 3.91                                                          & 5.84                                                     \\
					0.9      & 4.17                                                & 4.17                                                       &                   & 4.16                                                     & 1.19                                                       & 50.96                                               & 9.41       & 4.1                                                           & 6.29                                                     &                   & 2.93                                                     & 1.25                                                       & 51.08                                               & 9.55       & 4.2                                                           & 6.44                                                     \\
					1        & 4.46                                                & 4.46                                                       &                   & 4.46                                                     & 1                                                          & 50.91                                               & 10.27      & 4.4                                                           & 6.86                                                     &                   & 3.23                                                     & 1.13                                                       & 50.96                                               & 10.35      & 4.42                                                          & 6.95                                                                                                                                                
					\\       
					\hline
					
			\end{tabular}}
		}
		\label{tab1}
	\end{table}

	The portfolio is on $S_{t_1}$ when $\alpha=1,$  which is the one being traded in this case. In the  absence of transaction costs, the average simulated portfolio prices are closer to the unhedged value than when  costs are inclusive. The average simulated prices  excluding   transaction costs are higher than those with the inclusion of  costs for all values of $\alpha.$ In the presence of transaction costs, the average simulated prices have higher standard deviation values than those without transaction costs for all values of $\alpha \leq 0.5$ thus having a lower risk when the proportions of $S_{t_1}$ are the highest in the portfolio (the asset being traded). 
	
	Based on the standard deviation values and the simulated values with the  inclusion of transaction costs, hedging in this example is preferred over not trading. Comparing the simulated prices in the  presence and absence of transaction costs, the values are very high when $\alpha=0$ and 1 implying that, when the portfolio is on one of the assets, the expected return is higher than when it is blended.

	The Black-Scholes price is always equal to the unhedged value since  $r= \mu_1=\mu_2$. 
	The average simulated prices of the portfolio while trading $S_{t_1}$ are very close to the unhedged value when $\alpha=0$ and 1, and only differ owing to some noise in the simulations. The average simulated prices for the rest of the $\alpha$ values are smaller as they move away from $\alpha=0$ and 1 but also close to their respective theoretical unhedged values. That is, when the portfolio is blended, its  simulated portfolio prices are lower than  when the portfolio is only on one of the assets. Hedging more (252 times in this case) in the presence of transaction costs results in a high value of risk and  leads to a loss of money. The average simulated prices for $\alpha$ values that are $\neq 0$ or 1, are less than those for $\alpha=0$ and 1. The average simulated prices with the highest standard deviation are those close to $\alpha=0$, the one at zero with the highest value. It is the lowest at $\alpha=1,$ because  $S_{t_1}$  is the one being traded at this point. The difference in the discounted $P$ measure values was due to  noise in the simulations.

	\subsubsection{Case II: Trading only $S_{t_2}$;}\label{subsubsec:3.1.2}
\begin{figure}[htbp]
		
		\centering
		\subfloat[$k_2=0.0$.]{%
			\resizebox*{7cm}{!}{\includegraphics{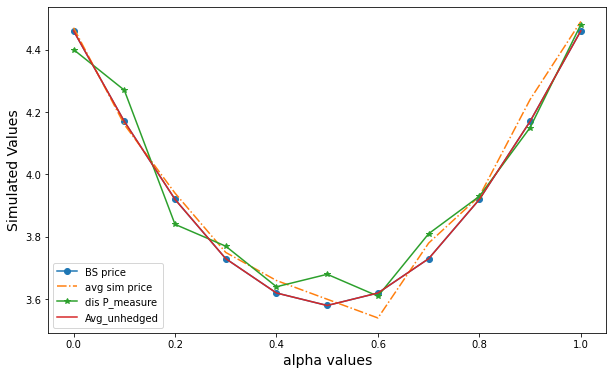}}}\hspace{5pt}
		\hspace{5pt}
		\subfloat[$k_2=0.001$.]{%
			\resizebox*{7cm}{!}{\includegraphics{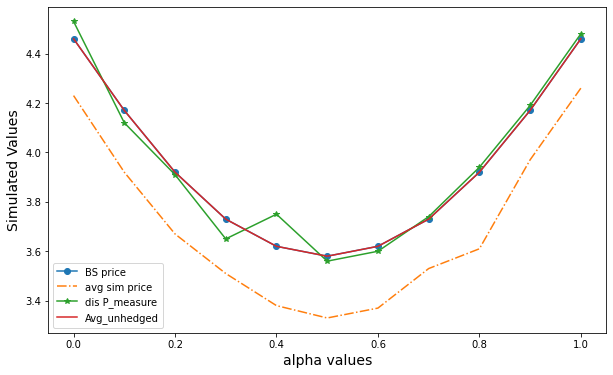}}}\hspace{5pt}
		
		\caption{ 
			$S0_1 = 50 ,~S0_2 = 50,~\rho = 0.2,~K = 50,~r = 0.01,~ T = 1 ,~\sigma_1 = 0.2, \sigma_2 = 0.2,~\mu_1 = 0.02~\mu_2 = 0.02,$
			num\_steps=252.} 
		\label{fig2}
	\end{figure}

	\vspace{0.15cm}
	\underline{Simulation I}~~
	In this simulation,  $r= \mu_1=\mu_2$ and  the parameters used are the following; $S0_1 = 50 ,~S0_2 = 50,~\rho = 0.2,~K = 50,~r = 0.02,~ T = 1 ,~\sigma_1 = 0.2, ~\sigma_2 = 0.2,~\mu_1 = 0.02,~\mu_2 = 0.02,$ num\_steps=252, 
	num\_simulations= 10000,~$k_2$=0.001, alpha\_values = 0 to 11. Leland's number  in this simulation:  A$<1.$ 
	The Portfolio is on $S_{t_2}$ when $\alpha=0.$ The average simulated prices of the portfolio while trading $S_{t_2}$ are high when $\alpha=0$ and 1. The average simulated prices for the rest of the $\alpha$ values are smaller as they move away from $\alpha=0$ and 1 but also close to their  theoretical unhedged value, especially in the  absence of transaction costs. The simulation results are presented in Table 
	\ref{tab2}  and Figure \ref{fig2}.

	\begin{table}[h]\tiny
		\caption{Without and with  transaction costs of  $S_{t_2}$.}
		{
			\resizebox{\columnwidth}{!}{%
				\begin{tabular}{ccccccccccccccccc}
					
					\hline
					\multicolumn{3}{c}{}                                                                                                        & \multirow{2}{*}{} & \multicolumn{6}{c}{\text{Without Transaction costs}}                                                                                                                                                                                                                                                                       & \multirow{2}{*}{} & \multicolumn{6}{c}{\text{With Transaction costs}}                                                                                                                                                                                                                                                                          \\  \cline{5-10} \cline{12-17} 
					\text{$\alpha$} & \begin{tabular}[c]{@{}c@{}}\text{BS}\\  \text{price}\end{tabular} & \begin{tabular}[c]{@{}c@{}}\text{Avg \_}\\   \text{ unhedged}\end{tabular} &                   & \begin{tabular}[c]{@{}c@{}}\text{avg sim} \\ \text{price}\end{tabular} & \begin{tabular}[c]{@{}c@{}}\text{std \_}\\ \text{sim price}\end{tabular} & \begin{tabular}[c]{@{}c@{}}\text{avg}\\  \text{S\_T}\end{tabular} & \text{std \_S\_T} & \begin{tabular}[c]{@{}c@{}}\text{dis P\_}\\ \text{mea}\\  \text{sure}\end{tabular} & \begin{tabular}[c]{@{}c@{}}\text{std \_}\\ \text{dis \_P}\end{tabular} &                   & \begin{tabular}[c]{@{}c@{}}\text{avg sim}\\  \text{price}\end{tabular} & \begin{tabular}[c]{@{}c@{}}\text{std \_}\\ \text{sim price}\end{tabular} & \begin{tabular}[c]{@{}c@{}}\text{avg}\\  \text{S\_T}\end{tabular} & \text{std \_S\_T }& \begin{tabular}[c]{@{}c@{}}\text{dis P\_}\\ \text{mea}\\  \text{sure}\end{tabular} & \begin{tabular}[c]{@{}c@{}}\text{std \_}\\ \text{dis \_P}\end{tabular} \\ \hline
					0        & 4.46     & 4.46            &  & 4.47                      & 1.02            & 50.92    & 10.23      & 4.4             & 6.8           &  & 4.23                  & 1.04            & 51.07    & 10.41      & 4.53            & 6.98          \\
					0.1      & 4.17     & 4.17            &  & 4.16                      & 1.2             & 51.2     & 9.55       & 4.27            & 6.45          &  & 3.92                  & 1.21            & 50.9     & 9.55       & 4.12            & 6.36          \\
					0.2      & 3.92     & 3.92            &  & 3.94                      & 1.85            & 50.85    & 8.82       & 3.84            & 5.85          &  & 3.67                  & 1.89            & 50.96    & 8.9        & 3.91            & 5.89          \\
					0.3      & 3.73     & 3.73            &  & 3.75                      & 2.76            & 51.04    & 8.46       & 3.77            & 5.66          &  & 3.51                  & 2.69            & 50.84    & 8.39       & 3.65            & 5.57          \\
					0.4      & 3.62     & 3.62            &  & 3.66                      & 3.57            & 51.07    & 8.06       & 3.64            & 5.4           &  & 3.38                  & 3.61            & 51.17    & 8.17       & 3.75            & 5.48          \\
					0.5      & 3.58     & 3.58            &  & 3.6                       & 4.64            & 51.16    & 8          & 3.68            & 5.33          &  & 3.33                  & 4.48            & 50.91    & 8.04       & 3.56            & 5.31          \\
					0.6      & 3.62     & 3.62            &  & 3.54                      & 5.51            & 51       & 8.07       & 3.61            & 5.37          &  & 3.37                  & 5.38            & 51.02    & 8.01       & 3.6             & 5.35          \\
					0.7      & 3.73     & 3.73            &  & 3.78                      & 6.41            & 51.07    & 8.52       & 3.81            & 5.72          &  & 3.53                  & 6.32            & 50.97    & 8.39       & 3.74            & 5.56          \\
					0.8      & 3.92     & 3.92            &  & 3.93                      & 7.22            & 51.05    & 8.92       & 3.93            & 5.99          &  & 3.61                  & 7.32            & 51.1     & 8.82       & 3.94            & 5.93          \\
					0.9      & 4.17     & 4.17            &  & 4.24                      & 7.84            & 50.95    & 9.53       & 4.15            & 6.31          &  & 3.97                  & 7.97            & 51.05    & 9.5        & 4.19            & 6.33          \\
					1        & 4.46     & 4.46            &  & 4.49                      & 8.78            & 51.03    & 10.3       & 4.48            & 6.88          &  & 4.26                  & 8.76            & 51       & 10.37      & 4.48            & 6.96         
					\\       
					\hline
					
				\end{tabular}
				
			}
		}
		\label{tab2}
	\end{table}

	In the presence of transaction costs, the average simulated prices have higher standard deviation values than  those without transaction costs, which decrease as $\alpha$ approaches zero. This is because the asset being traded in this case is the second asset, and our portfolio is  only on $S_{t_2}$ which is the one being traded, hence having a lower risk when the proportion of $S_{t_2}$ is the highest in the portfolio. As in Case I, the average simulated prices are the highest when $\alpha=0$ and 1. However, in all cases, we are better off  trading than not trading based on the standard deviation values of the discounted $P$ measure value and those of the average simulated price.

	In conclusion, while trading only one  of the assets in the whole portfolio, the average simulated portfolio prices in a simulation are the highest at the edges of the $\alpha$ values, that is, when $\alpha=0$ and 1. More money is lost with the inclusion of transaction costs than with hedging,  excluding transaction costs.
	In the next section, we use $RAV$ to draw conclusions on what to trade rather than based on standard deviation values alone.

	\subsection{Hedging one of the assets when the portfolio is on one of them}\label{subsec:4.2}
	In this subsection, we aim to determine  when it is better to trade the wrong asset, right asset, or not trading at all in our portfolio. We conduct simulations in which we have a case for trading only asset 1 (as the one done in subsubsection \ref{subsubsec:3.1.1}) and another case when we can only trade asset 2 (as the one done in subsubsection \ref{subsubsec:3.1.2}). Different $\rho$ values and  transaction costs were used for each asset simultaneously. Different rebalancing intervals were considered to aid in answering the research question. 
	
	We hedge only one asset in our portfolio, and from the conclusion in subsubsections \ref{subsubsec:3.1.1} and \ref{subsubsec:3.1.2}, 
	while trading only one of the assets in the whole portfolio, the average simulated portfolio prices are highest at the edges of the $\alpha$ values, that is, when $\alpha=0$ and 1. While trading $S_{t_1}$ only, $\alpha=0$ and 1  have average simulated portfolio prices  larger than those when $\alpha\neq 0$ and $\alpha \neq 1$. This is the  case when trading only $S_{t_2}$ in the same market scenario. 
	
	Therefore, we aim to determine when it is possible to trade the wrong asset when $\alpha=$ 0 and 1 (when the portfolio is on one of the assets). We consider the wrong asset to be $S_{t_2}$ when $\alpha=1$ and the right asset to be $S_{t_1}.$  When $\alpha=0,$ we consider the wrong asset to be $S_{t_1}$ and the right one as $S_{t_2}.$
	For the other values of $\alpha$, the simulated portfolio prices are expected to be between those of $\alpha=0$ and 1 (they increase as they approach 0 and 1), as observed in the previous subsection.  Running multiple simulations generates more statistical and relevant results that aid in estimating the randomness of the simulated values, hence being aware of  the risk incurred. This ensured that outliers did not affect the reliability of the  conclusions obtained from the simulated results.  In all  experiments, the number of simulations were 10000. The parameters used were as previously described.
	%
	%
	
	The results are presented in tables and the following symbols were utilized:
	WA, RA and NA represent wrong asset, right asset, and no assets traded at all, respectively.
	
	$\checkmark\checkmark$  represents the best decision to opt for, 
	
	$\checkmark$ represents the second best decision to opt for and 
	
	$x$ represents the worst decision to undertake.
	\vspace{0.2cm}
	\begin{conjecture}
		With the inclusion of transaction costs, there is a value of correlation above which we can not trade the right asset but trade the wrong asset.
	\end{conjecture}
	\vspace{0.1cm}
	In the same simulation code, we carried out Monte Carlo simulations for the  simulated portfolio price,  standard deviation of the simulated portfolio price, unhedged value, and the discounted $P$ measure value while trading each of the assets alone in the  presence and absence of transaction costs.
	When $\alpha=0,$ the portfolio is on $S_{t_1}$ and in this case, the right asset to be traded in the portfolio is the first asset (wrong asset is $S_{t_2}$),  and on $S_{t_2}$ when $\alpha=1$ hence, the right asset to be traded in this case is the second asset (wrong asset is $S_{t_1}$).
	Conclusions on when to trade the right asset, wrong asset, or not to trade at all in each case were made based on the $RAV$ of the average simulated portfolio prices in each case and the discounted $P$ measure values in the presence and absence of transaction costs.

	\subsubsection{Decisions Based on the Risk Adjusted Value ($RAV$)}
	The $RAV$ considered in all  simulations in this study uses the market price of risk $\lambda$, and the best decision to undertake from trading the right asset, wrong asset, or not trading is the one with the highest $RAV.$ For all our simulations $\lambda$ is assumed to be equal to different values; 2,0.5,0.3 and 0.2 in all our simulations, to help in drawing meaningful conclusions when returns and volatility are high or low. When $\lambda$ is small, an investor does not care about  risk.
	
	We considered the same initial parameters for both  assets; hence, the conclusions when $\alpha=0$ are not different from those when $\alpha=1$ but  the right and wrong assets change. Therefore, in our simulations, the calculations are for $\alpha=0,$ number of simulations = 10000, ~
	number of steps $= [2,4, 12, 52, 252],$ $T=1$~ and 
	$\rho$ = $[0.2,0.3,0.35,0.5,0.55,0.7,0.8,0.99]$. In the results presented in the tables, $k_{\text{right}}$ represents $k_2$ because we  have the portfolio on $S_{t_2}$ when $\alpha=0 $ and $k_{\text{wrong}}$ represents $k_1$ hence, the wrong asset is  $S_{t_1}$ and the right one is $S_{t_2}.$

	\subsubsection{Transaction Costs of 0.1\% on the right asset and 0\% on the wrong asset}
	First, we consider zero transaction costs on the wrong asset and 0.001(very small) transaction costs on the right asset.
	All Leland's numbers  in this simulation are less than one for all rebalancing intervals because the transaction costs are very small. In this simulation, the value of $\alpha=0$ hence, the right asset to be traded is the second one. 
	Different values of $\lambda$ were considered in the computations of the $RAV$  and the $\alpha$ value considered throughout was 0. The discounted $P$ measure value and its standard deviation are utilized in the computation of the $RAV$ for trading none of the assets.  The volatility of the two assets was considered to be the same in the simulation.

	The transaction costs of 0.1\%  were the smallest  considered in our simulations. The conclusions of the $RAV$ are listed in Table~\ref{tab3} for $\rho=0.99.$ Obtaining the $RAV$ using a higher value of $\lambda= 2.0,$ the best decision to undertake is to trade the right asset in the presence and absence of transaction costs for any $\rho$ value and  rebalancing interval.

	\begin{table}[h] \centering
		\caption{Some of the conclusions from $RAV$ values with $S0_1 = S0_2 = K = 50,~r = 0.02,~\sigma_1 =\sigma_2 = 0.2,~
			\mu_1 = \mu_2 = 0.05,$ $\alpha=0,~k_{\text{wrong}}=0.0,~k_{\text{right}}=0.001.$}
		{	\resizebox{!}{1.02cm}{		
				\begin{tabular}{ccccccccccccccccccccccccccccc}				\hline
					
					\multicolumn{2}{c}{\multirow{3}{*}{}}                                         & \multicolumn{6}{c}{ \text{$\lambda=2$}}                                                                               & \multirow{4}{*}{} & \multicolumn{6}{c}{\text{$\lambda=0.5$}}                                                                               & \multirow{4}{*}{} & \multicolumn{6}{c}{\text{$\lambda=0.3$}}                                                                               & \multirow{4}{*}{} & \multicolumn{6}{c}{\text{$\lambda=0.2$}}                                                                               \\ \cline{3-8} \cline{10-15} \cline{17-22} \cline{24-29} 
					\multicolumn{2}{c}{}                                                          & \multicolumn{3}{c}{\multirow{2}{*}{\begin{tabular}[c]{@{}c@{}}\text{With}\\ \text{Transaction Costs}\end{tabular}}} & \multicolumn{3}{c}{\multirow{2}{*}{\begin{tabular}[c]{@{}c@{}}\text{Without}\\ \text{Transaction Costs}\end{tabular}}} &                   & \multicolumn{3}{c}{\multirow{2}{*}{\begin{tabular}[c]{@{}c@{}}\text{With} \\ \text{Transaction Costs}\end{tabular}}} & \multicolumn{3}{c}{\multirow{2}{*}{\begin{tabular}[c]{@{}c@{}}\text{Without} \\ \text{Transaction Costs}\end{tabular}}} &                   & \multicolumn{3}{c}{\multirow{2}{*}{\begin{tabular}[c]{@{}c@{}}\text{With} \\ \text{Transaction Costs}\end{tabular}}} & \multicolumn{3}{c}{\multirow{2}{*}{\begin{tabular}[c]{@{}c@{}}\text{Without} \\ \text{Transaction Costs}\end{tabular}}} &                   & \multicolumn{3}{c}{\multirow{2}{*}{\begin{tabular}[c]{@{}c@{}}With \\ \text{Transaction} \text{Costs}\end{tabular}}} & \multicolumn{3}{c}{\multirow{2}{*}{\begin{tabular}[c]{@{}c@{}}\text{Without} \\ \text{Transaction Costs}\end{tabular}}} \\
					\multicolumn{2}{c}{}                                                          & \multicolumn{3}{c}{}                                                                                  & \multicolumn{3}{c}{}                                                                                     &                   & \multicolumn{3}{c}{}                                                                                   & \multicolumn{3}{c}{}                                                                                      &                   & \multicolumn{3}{c}{}                                                                                   & \multicolumn{3}{c}{}                                                                                      &                   & \multicolumn{3}{c}{}                                                                                   & \multicolumn{3}{c}{}                                                                                      \\ \cline{1-8} \cline{10-15} \cline{17-22} \cline{24-29} 
					\text{$\rho$ }               & \begin{tabular}[c]{@{}c@{}}\text{num\_}\\ \text{steps}\end{tabular} & \text{WA}                            & \text{RA}                                      & \text{NA}                          & \text{WA}                            & \text{RA}                                        & \text{NA}                           &                   & \text{WA}                           & \text{RA}                                       & \text{NA}                           & \text{WA}                            & \text{RA}                                        & \text{NA}                            &                   & \text{WA}                                   & \text{RA}                                   & \text{NA}                       & \text{WA}                            & \text{RA}                                        & \text{NA}                            &                   & \text{WA}                               & \text{RA}                               & \text{NA}                               & \text{WA}                            & \text{RA}                                        & \text{NA}                          \\\hline
					\multirow{5}{*}{0.99} & 2                                                     & $\checkmark$                  & $\checkmark$$\checkmark$                & $x$                         & $\checkmark$                  & $\checkmark$$\checkmark$                  & $x$                          & \multirow{5}{*}{} & $\checkmark$                         & $\checkmark$$\checkmark$             & $x$                      & $\checkmark$                  & $\checkmark$$\checkmark$                  & $x$                           & \multirow{5}{*}{} & $\checkmark$                         & $\checkmark$$\checkmark$             & $x$                      & $\checkmark$                  & $\checkmark$$\checkmark$                  & $x$                           & \multirow{5}{*}{} & $\checkmark$$\checkmark$             & $\checkmark$                         & $x$                      & $\checkmark$                  & $\checkmark$$\checkmark$                  & $x$                           \\
					& 4                                                     & $\checkmark$                  & $\checkmark$$\checkmark$                & $x$                         & $\checkmark$                  & $\checkmark$$\checkmark$                  & $x$                          &                   & $\checkmark$                         & $\checkmark$$\checkmark$             & $x$                      & $\checkmark$                  & $\checkmark$$\checkmark$                  & $x$                           &                   & $\checkmark$                         & $\checkmark$$\checkmark$             & $x$                      & $\checkmark$                  & $\checkmark$$\checkmark$                  & $x$                           &                   & $\checkmark$$\checkmark$             & $\checkmark$                         & $x$                      & $\checkmark$                  & $\checkmark$$\checkmark$                  & $x$                           \\
					& 12                                                    & $\checkmark$                  & $\checkmark$$\checkmark$                & $x$                         & $\checkmark$                  & $\checkmark$$\checkmark$                  & $x$                          &                   & $\checkmark$                         & $\checkmark$$\checkmark$             & $x$                      & $\checkmark$                  & $\checkmark$$\checkmark$                  & $x$                           &                   & $\checkmark$                         & $\checkmark$$\checkmark$             & $x$                      & $\checkmark$                  & $\checkmark$$\checkmark$                  & $x$                           &                   & $\checkmark$$\checkmark$             & $\checkmark$                         & $x$                      & $\checkmark$                  & $\checkmark$$\checkmark$                  & $x$                           \\
					& 52                                                    & $\checkmark$                  & $\checkmark$$\checkmark$                & $x$                         & $\checkmark$                  & $\checkmark$$\checkmark$                  & $x$                          &                   & $\checkmark$                         & $\checkmark$$\checkmark$             & $x$                      & $\checkmark$                  & $\checkmark$$\checkmark$                  & $x$                           &                   & $\checkmark$                         & $\checkmark$$\checkmark$             & $x$                      & $\checkmark$                  & $\checkmark$$\checkmark$                  & $x$                           &                   & $\checkmark$$\checkmark$             & $\checkmark$                         & $x$                      & $\checkmark$                  & $\checkmark$$\checkmark$                  & $x$                           \\
					& 252                                                   & $\checkmark$                  & $\checkmark$$\checkmark$                & $x$                         & $\checkmark$                  & $\checkmark$$\checkmark$                  & $x$                          &                   & $\checkmark$$\checkmark$             & $\checkmark$                         & $x$                      & $\checkmark$                  & $\checkmark$$\checkmark$                  & $x$                           &                   & $\checkmark$$\checkmark$             & $\checkmark$                         & $x$                      & $\checkmark$                  & $\checkmark$$\checkmark$                  & $x$                           &                   & $\checkmark$$\checkmark$             & $\checkmark$                         & $x$                      & $\checkmark$                  & $\checkmark$$\checkmark$                  & $x$                          
					\\       
					\hline
					
				\end{tabular}
				
		}}
		\label{tab3}
	\end{table}
	
	It is worth noting that even a highly correlated pair of stocks $(\rho = 0.99)$ still has a 
	large component of uncorrelated risk. Recall that if $Z_1$ and $Z_2$ are $\mathcal{N}(0, 1)$  variables with
	correlation $\rho$, we can simulate $Z_2 = \rho Z_1 + \sqrt{(1-\rho^2)}~W,$ where W $\sim \mathcal{N}(0, 1)$
	is uncorrelated with either $Z_1$ or $Z_2.$ Inserting $\rho = 0.99$ above reveals $Z_2 = 0.99~Z_1
	+ 0.141~W$. This simple calculation is a helpful reminder of  how big rho should be,
	before the two assets move in the true lockstep.
	
	Although the transaction costs on the wrong asset are zero, in this case trading the right asset is the best decision compared to trading the wrong asset because the transaction costs on the right asset are too small $(0.001),$  although the market price of risk is high. In this case, to compare the decision of trading the wrong asset and not trading at all, it is better not to hedge at all than trading the wrong asset when $\rho$ is  $0.2,0.3,0.35,0.5$ and 0.55 in the presence and absence of transaction costs, while when $\rho$ is high that is; $0.7,0.8$ and $0.99,$ it is better to trade the wrong asset than not to hedge at all in the presence and absence of transaction costs. 
	
	When $\lambda$ decreases to $0.5,$ the best decision remains to trade the right asset in presence and absence of transaction costs for any $\rho$ value and all rebalancing intervals except when $\rho=0.99$ for 252 rebalancing intervals where trading the wrong asset is the best decision. When $\rho \leq 0.7,$ no hedging  is better than trading the wrong asset, and when $\rho> 0.7,$ trading the wrong asset is better than not  trading at all in the  presence and absence of transaction costs.

	When $\rho=0.99$ and  $\lambda=0.3,$ trading the wrong asset  is the best decision at 252 rebalancing intervals in a year, and at other rebalancing intervals, trading the right asset wins, with the worst decision as not hedging at all.  Trading the right asset is the best decision for other $\rho$ values. 
	When $\rho=0.99$ and $\lambda=0.2,$ the best decision is to trade the wrong asset for all steps, and  the worst decision is always not to hedge at all. Trading the right asset is the best decision for other $\rho$ values.  Because the transaction costs on the right asset are also very low $(\approx 0),$ having a very high value of $\lambda$ ($\lambda=2$) does not favor the trading of the wrong asset at any $\rho$ value and at any rebalancing interval. Therefore, for a very high $\lambda$ where one would care about  risk, it is better to trade the right asset than the wrong one at $\rho=0.99$ when the transaction costs on the right asset are too small.

	\subsubsection{ Transaction Costs of 1\% on the right asset and 0\% on the wrong asset}

	In the next simulation, the transaction costs on the wrong asset are zero, whereas those on the right asset are 1\% for different $\lambda$ values. 

	For a  1\% transaction costs on the right asset and zero on the wrong asset, trading the wrong asset is the best decision when $\rho=0.99.$ The best decision, in the absence of transaction costs, is to trade the right asset for all $\rho$ and $\lambda$ at all rebalancing intervals. When $\lambda=2$ in the  presence of (reasonable) transaction costs (on the right asset), the best decision is to trade the right asset for all $\rho \leq 0.8.$ Not hedging is better than trading the wrong asset for all $\rho \leq 0.55,$ and  for $\rho \geq 0.7$ not hedging is the worst option. A summary of the conclusions for all the $\rho$ values is presented in Table \ref{tab4}.

	\begin{table}[h]\tiny \centering
		\caption{Conclusions from the calculated $RAV$ values when $S0_1 = S0_2 = K = 50,~r = 0.02,~\sigma_1 =\sigma_2 = 0.2,~
			\mu_1 = \mu_2 = 0.05,~k_{\text{wrong}}=0,~k_{\text{right}}=0.01.$                
		}
		{ \resizebox{!}{4.2cm}{
				\begin{tabular}{ccccccccccccccccccccccccccccc}
					\hline
					\multicolumn{2}{c}{\multirow{3}{*}{}}                                         & \multicolumn{6}{c}{ \text{$\lambda=2$}}                                                                               & \multirow{4}{*}{} & \multicolumn{6}{c}{\text{$\lambda=0.5$}}                                                                               & \multirow{4}{*}{} & \multicolumn{6}{c}{\text{$\lambda=0.3$}}                                                                               & \multirow{4}{*}{} & \multicolumn{6}{c}{\text{$\lambda=0.2$}}                                                                               \\ \cline{3-8} \cline{10-15} \cline{17-22} \cline{24-29} 
					\multicolumn{2}{c}{}                                                          & \multicolumn{3}{c}{\multirow{2}{*}{\begin{tabular}[c]{@{}c@{}}\text{With}\\ \text{Transaction Costs}\end{tabular}}} & \multicolumn{3}{c}{\multirow{2}{*}{\begin{tabular}[c]{@{}c@{}}\text{Without}\\ \text{Transaction Costs}\end{tabular}}} &                   & \multicolumn{3}{c}{\multirow{2}{*}{\begin{tabular}[c]{@{}c@{}}\text{With} \\ \text{Transaction Costs}\end{tabular}}} & \multicolumn{3}{c}{\multirow{2}{*}{\begin{tabular}[c]{@{}c@{}}\text{Without} \\ \text{Transaction Costs}\end{tabular}}} &                   & \multicolumn{3}{c}{\multirow{2}{*}{\begin{tabular}[c]{@{}c@{}}\text{With} \\ \text{Transaction Costs}\end{tabular}}} & \multicolumn{3}{c}{\multirow{2}{*}{\begin{tabular}[c]{@{}c@{}}\text{Without} \\ \text{Transaction Costs}\end{tabular}}} &                   & \multicolumn{3}{c}{\multirow{2}{*}{\begin{tabular}[c]{@{}c@{}}With \\ \text{Transaction} \text{Costs}\end{tabular}}} & \multicolumn{3}{c}{\multirow{2}{*}{\begin{tabular}[c]{@{}c@{}}\text{Without} \\ \text{Transaction Costs}\end{tabular}}} \\
					\multicolumn{2}{c}{}                                                          & \multicolumn{3}{c}{}                                                                                  & \multicolumn{3}{c}{}                                                                                     &                   & \multicolumn{3}{c}{}                                                                                   & \multicolumn{3}{c}{}                                                                                      &                   & \multicolumn{3}{c}{}                                                                                   & \multicolumn{3}{c}{}                                                                                      &                   & \multicolumn{3}{c}{}                                                                                   & \multicolumn{3}{c}{}                                                                                      \\ \cline{1-8} \cline{10-15} \cline{17-22} \cline{24-29} 
					\text{$\rho$ }               & \begin{tabular}[c]{@{}c@{}}\text{num\_}\\ \text{steps}\end{tabular} & \text{WA}                            & \text{RA}                                      & \text{NA}                          & \text{WA}                            & \text{RA}                                        & \text{NA}                           &                   & \text{WA}                           & \text{RA}                                       & \text{NA}                           & \text{WA}                            & \text{RA}                                        & \text{NA}                            &                   & \text{WA}                                   & \text{RA}                                   & \text{NA}                       & \text{WA}                            & \text{RA}                                        & \text{NA}                            &                   & \text{WA}                               & \text{RA}                               & \text{NA}                               & \text{WA}                            & \text{RA}                                        & \text{NA}                          \\\hline		\multirow{5}{*}{0.2}  & 2                                                     & $x$                                             & $\checkmark$$\checkmark$           & $\checkmark$   & $x$                           & $\checkmark$$\checkmark$                  & $\checkmark$                 & \multirow{5}{*}{} & $x$                              & $\checkmark$$\checkmark$         & $\checkmark$                     & $x$                           & $\checkmark$$\checkmark$                  & $\checkmark$                  & \multirow{5}{*}{} & $x$                              & $\checkmark$$\checkmark$         & $\checkmark$                     & $x$                           & $\checkmark$$\checkmark$                  & $\checkmark$                  & \multirow{5}{*}{} & $x$                                   & $\checkmark$             & $\checkmark$$\checkmark$            & $x$                           & $\checkmark$$\checkmark$                  & $\checkmark$                  \\
					& 4                                                     & $x$                                             & $\checkmark$$\checkmark$           & $\checkmark$   & $x$                           & $\checkmark$$\checkmark$                  & $\checkmark$                 &                   & $x$                              & $\checkmark$$\checkmark$         & $\checkmark$                     & $x$                           & $\checkmark$$\checkmark$                  & $\checkmark$                  &                   & $x$                              & $\checkmark$$\checkmark$         & $\checkmark$                     & $x$                           & $\checkmark$$\checkmark$                  & $\checkmark$                  &                   & $x$                                   & $\checkmark$             & $\checkmark$$\checkmark$            & $x$                           & $\checkmark$$\checkmark$                  & $\checkmark$                  \\
					& 12                                                    & $x$                                             & $\checkmark$$\checkmark$           & $\checkmark$   & $x$                           & $\checkmark$$\checkmark$                  & $\checkmark$                 &                   & $x$                              & $\checkmark$$\checkmark$         & $\checkmark$                     & $x$                           & $\checkmark$$\checkmark$                  & $\checkmark$                  &                   & $x$                              & $\checkmark$$\checkmark$         & $\checkmark$                     & $x$                           & $\checkmark$$\checkmark$                  & $\checkmark$                  &                   & $x$                                   & $\checkmark$             & $\checkmark$$\checkmark$            & $x$                           & $\checkmark$$\checkmark$                  & $\checkmark$                  \\
					& 52                                                    & $x$                                             & $\checkmark$$\checkmark$           & $\checkmark$   & $x$                           & $\checkmark$$\checkmark$                  & $\checkmark$                 &                   & $x$                              & $\checkmark$$\checkmark$         & $\checkmark$                     & $x$                           & $\checkmark$$\checkmark$                  & $\checkmark$                  &                   & $x$                              & $\checkmark$                     & $\checkmark$$\checkmark$         & $x$                           & $\checkmark$$\checkmark$                  & $\checkmark$                  &                   & $x$                                   & $\checkmark$             & $\checkmark$$\checkmark$            & $x$                           & $\checkmark$$\checkmark$                  & $\checkmark$                  \\
					& 252                                                   & $x$                                             & $\checkmark$$\checkmark$           & $\checkmark$   & $x$                           & $\checkmark$$\checkmark$                  & $\checkmark$                 &                   & $x$                              & $\checkmark$                     & $\checkmark$$\checkmark$         & $x$                           & $\checkmark$$\checkmark$                  & $\checkmark$                  &                   & $x$                              & $\checkmark$                     & $\checkmark$$\checkmark$         & $x$                           & $\checkmark$$\checkmark$                  & $\checkmark$                  &                   & $\checkmark$                          & $x$                      & $\checkmark$$\checkmark$            & $x$                           & $\checkmark$$\checkmark$                  & $\checkmark$                  \\\hline
					\multirow{5}{*}{0.3}  & 2                                                     & $x$                                             & $\checkmark$$\checkmark$           & $\checkmark$   & $x$                           & $\checkmark$$\checkmark$                  & $\checkmark$                 & \multirow{5}{*}{} & $x$                              & $\checkmark$$\checkmark$         & $\checkmark$                     & $x$                           & $\checkmark$$\checkmark$                  & $\checkmark$                  & \multirow{5}{*}{} & $x$                              & $\checkmark$$\checkmark$         & $\checkmark$                     & $x$                           & $\checkmark$$\checkmark$                  & $\checkmark$                  & \multirow{5}{*}{} & $x$                                   & $\checkmark$             & $\checkmark$$\checkmark$            & $x$                           & $\checkmark$$\checkmark$                  & $\checkmark$                  \\
					& 4                                                     & $x$                                             & $\checkmark$$\checkmark$           & $\checkmark$   & $x$                           & $\checkmark$$\checkmark$                  & $\checkmark$                 &                   & $x$                              & $\checkmark$$\checkmark$         & $\checkmark$                     & $x$                           & $\checkmark$$\checkmark$                  & $\checkmark$                  &                   & $x$                              & $\checkmark$$\checkmark$         & $\checkmark$                     & $x$                           & $\checkmark$$\checkmark$                  & $\checkmark$                  &                   & $x$                                   & $\checkmark$             & $\checkmark$$\checkmark$            & $x$                           & $\checkmark$$\checkmark$                  & $\checkmark$                  \\
					& 12                                                    & $x$                                             & $\checkmark$$\checkmark$           & $\checkmark$   & $x$                           & $\checkmark$$\checkmark$                  & $\checkmark$                 &                   & $x$                              & $\checkmark$$\checkmark$         & $\checkmark$                     & $x$                           & $\checkmark$$\checkmark$                  & $\checkmark$                  &                   & $x$                              & $\checkmark$$\checkmark$         & $\checkmark$                     & $x$                           & $\checkmark$$\checkmark$                  & $\checkmark$                  &                   & $x$                                   & $\checkmark$             & $\checkmark$$\checkmark$            & $x$                           & $\checkmark$$\checkmark$                  & $\checkmark$                  \\
					& 52                                                    & $x$                                             & $\checkmark$$\checkmark$           & $\checkmark$   & $x$                           & $\checkmark$$\checkmark$                  & $\checkmark$                 &                   & $x$                              & $\checkmark$$\checkmark$         & $\checkmark$                     & $x$                           & $\checkmark$$\checkmark$                  & $\checkmark$                  &                   & $x$                              & $\checkmark$                     & $\checkmark$$\checkmark$         & $x$                           & $\checkmark$$\checkmark$                  & $\checkmark$                  &                   & $x$                                   & $\checkmark$             & $\checkmark$$\checkmark$            & $x$                           & $\checkmark$$\checkmark$                  & $\checkmark$                  \\
					& 252                                                   & $x$                                             & $\checkmark$$\checkmark$           & $\checkmark$   & $x$                           & $\checkmark$$\checkmark$                  & $\checkmark$                 &                   & $x$                              & $\checkmark$                     & $\checkmark$$\checkmark$         & $x$                           & $\checkmark$$\checkmark$                  & $\checkmark$                  &                   & $\checkmark$                     & $x$                              & $\checkmark$$\checkmark$         & $x$                           & $\checkmark$$\checkmark$                  & $\checkmark$                  &                   & $\checkmark$                          & $x$                      & $\checkmark$$\checkmark$            & $x$                           & $\checkmark$$\checkmark$                  & $\checkmark$                  \\\hline
					\multirow{5}{*}{0.35} & 2                                                     & $x$                                             & $\checkmark$$\checkmark$           & $\checkmark$   & $x$                           & $\checkmark$$\checkmark$                  & $\checkmark$                 & \multirow{5}{*}{} & $x$                              & $\checkmark$$\checkmark$         & $\checkmark$                     & $x$                           & $\checkmark$$\checkmark$                  & $\checkmark$                  & \multirow{5}{*}{} & $x$                              & $\checkmark$$\checkmark$         & $\checkmark$                     & $x$                           & $\checkmark$$\checkmark$                  & $\checkmark$                  & \multirow{5}{*}{} & $x$                                   & $\checkmark$             & $\checkmark$$\checkmark$            & $x$                           & $\checkmark$$\checkmark$                  & $\checkmark$                  \\
					& 4                                                     & $x$                                             & $\checkmark$$\checkmark$           & $\checkmark$   & $x$                           & $\checkmark$$\checkmark$                  & $\checkmark$                 &                   & $x$                              & $\checkmark$$\checkmark$         & $\checkmark$                     & $x$                           & $\checkmark$$\checkmark$                  & $\checkmark$                  &                   & $x$                              & $\checkmark$$\checkmark$         & $\checkmark$                     & $x$                           & $\checkmark$$\checkmark$                  & $\checkmark$                  &                   & $x$                                   & $\checkmark$             & $\checkmark$$\checkmark$            & $x$                           & $\checkmark$$\checkmark$                  & $\checkmark$                  \\
					& 12                                                    & $x$                                             & $\checkmark$$\checkmark$           & $\checkmark$   & $x$                           & $\checkmark$$\checkmark$                  & $\checkmark$                 &                   & $x$                              & $\checkmark$$\checkmark$         & $\checkmark$                     & $x$                           & $\checkmark$$\checkmark$                  & $\checkmark$                  &                   & $x$                              & $\checkmark$$\checkmark$         & $\checkmark$                     & $x$                           & $\checkmark$$\checkmark$                  & $\checkmark$                  &                   & $x$                                   & $\checkmark$             & $\checkmark$$\checkmark$            & $x$                           & $\checkmark$$\checkmark$                  & $\checkmark$                  \\
					& 52                                                    & $x$                                             & $\checkmark$$\checkmark$           & $\checkmark$   & $x$                           & $\checkmark$$\checkmark$                  & $\checkmark$                 &                   & $x$                              & $\checkmark$$\checkmark$         & $\checkmark$                     & $x$                           & $\checkmark$$\checkmark$                  & $\checkmark$                  &                   & $x$                              & $\checkmark$                     & $\checkmark$$\checkmark$         & $x$                           & $\checkmark$$\checkmark$                  & $\checkmark$                  &                   & $x$                                   & $\checkmark$             & $\checkmark$$\checkmark$            & $x$                           & $\checkmark$$\checkmark$                  & $\checkmark$                  \\
					& 252                                                   & $x$                                             & $\checkmark$$\checkmark$           & $\checkmark$   & $x$                           & $\checkmark$$\checkmark$                  & $\checkmark$                 &                   & $x$                              & $\checkmark$                     & $\checkmark$$\checkmark$         & $x$                           & $\checkmark$$\checkmark$                  & $\checkmark$                  &                   & $\checkmark$                     & $x$                              & $\checkmark$$\checkmark$         & $x$                           & $\checkmark$$\checkmark$                  & $\checkmark$                  &                   & $\checkmark$                          & $x$                      & $\checkmark$$\checkmark$            & $x$                           & $\checkmark$$\checkmark$                  & $\checkmark$                  \\\hline
					\multirow{5}{*}{0.5}  & 2                                                     & $x$                                             & $\checkmark$$\checkmark$           & $\checkmark$   & $x$                           & $\checkmark$$\checkmark$                  & $\checkmark$                 & \multirow{5}{*}{} & $x$                              & $\checkmark$$\checkmark$         & $\checkmark$                     & $x$                           & $\checkmark$$\checkmark$                  & $\checkmark$                  & \multirow{5}{*}{} & $x$                              & $\checkmark$$\checkmark$         & $\checkmark$                     & $x$                           & $\checkmark$$\checkmark$                  & $\checkmark$                  & \multirow{5}{*}{} & $x$                                   & $\checkmark$             & $\checkmark$$\checkmark$            & $x$                           & $\checkmark$$\checkmark$                  & $\checkmark$                  \\
					& 4                                                     & $x$                                             & $\checkmark$$\checkmark$           & $\checkmark$   & $x$                           & $\checkmark$$\checkmark$                  & $\checkmark$                 &                   & $x$                              & $\checkmark$$\checkmark$         & $\checkmark$                     & $x$                           & $\checkmark$$\checkmark$                  & $\checkmark$                  &                   & $x$                              & $\checkmark$$\checkmark$         & $\checkmark$                     & $x$                           & $\checkmark$$\checkmark$                  & $\checkmark$                  &                   & $x$                                   & $\checkmark$             & $\checkmark$$\checkmark$            & $x$                           & $\checkmark$$\checkmark$                  & $\checkmark$                  \\
					& 12                                                    & $x$                                             & $\checkmark$$\checkmark$           & $\checkmark$   & $x$                           & $\checkmark$$\checkmark$                  & $\checkmark$                 &                   & $x$                              & $\checkmark$$\checkmark$         & $\checkmark$                     & $x$                           & $\checkmark$$\checkmark$                  & $\checkmark$                  &                   & $x$                              & $\checkmark$$\checkmark$         & $\checkmark$                     & $x$                           & $\checkmark$$\checkmark$                  & $\checkmark$                  &                   & $x$                                   & $\checkmark$             & $\checkmark$$\checkmark$            & $x$                           & $\checkmark$$\checkmark$                  & $\checkmark$                  \\
					& 52                                                    & $x$                                             & $\checkmark$$\checkmark$           & $\checkmark$   & $x$                           & $\checkmark$$\checkmark$                  & $\checkmark$                 &                   & $x$                              & $\checkmark$$\checkmark$         & $\checkmark$                     & $x$                           & $\checkmark$$\checkmark$                  & $\checkmark$                  &                   & $x$                              & $\checkmark$                     & $\checkmark$$\checkmark$         & $x$                           & $\checkmark$$\checkmark$                  & $\checkmark$                  &                   & $x$                                   & $\checkmark$             & $\checkmark$$\checkmark$            & $x$                           & $\checkmark$$\checkmark$                  & $\checkmark$                  \\
					& 252                                                   & $x$                                             & $\checkmark$$\checkmark$           & $\checkmark$   & $x$                           & $\checkmark$$\checkmark$                  & $\checkmark$                 &                   & $x$                              & $\checkmark$                     & $\checkmark$$\checkmark$         & $x$                           & $\checkmark$$\checkmark$                  & $\checkmark$                  &                   & $\checkmark$                     & $x$                              & $\checkmark$$\checkmark$         & $x$                           & $\checkmark$$\checkmark$                  & $\checkmark$                  &                   & $\checkmark$                          & $x$                      & $\checkmark$$\checkmark$            & $x$                           & $\checkmark$$\checkmark$                  & $\checkmark$                  \\\hline
					\multirow{5}{*}{0.55} & 2                                                     & $x$                                             & $\checkmark$$\checkmark$           & $\checkmark$   & $x$                           & $\checkmark$$\checkmark$                  & $\checkmark$                 & \multirow{5}{*}{} & $x$                              & $\checkmark$$\checkmark$         & $\checkmark$                     & $x$                           & $\checkmark$$\checkmark$                  & $\checkmark$                  & \multirow{5}{*}{} & $x$                              & $\checkmark$$\checkmark$         & $\checkmark$                     & $x$                           & $\checkmark$$\checkmark$                  & $\checkmark$                  & \multirow{5}{*}{} & $x$                                   & $\checkmark$             & $\checkmark$$\checkmark$            & $x$                           & $\checkmark$$\checkmark$                  & $\checkmark$                  \\
					& 4                                                     & $x$                                             & $\checkmark$$\checkmark$           & $\checkmark$   & $x$                           & $\checkmark$$\checkmark$                  & $\checkmark$                 &                   & $x$                              & $\checkmark$$\checkmark$         & $\checkmark$                     & $x$                           & $\checkmark$$\checkmark$                  & $\checkmark$                  &                   & $x$                              & $\checkmark$$\checkmark$         & $\checkmark$                     & $x$                           & $\checkmark$$\checkmark$                  & $\checkmark$                  &                   & $x$                                   & $\checkmark$             & $\checkmark$$\checkmark$            & $x$                           & $\checkmark$$\checkmark$                  & $\checkmark$                  \\
					& 12                                                    & $x$                                             & $\checkmark$$\checkmark$           & $\checkmark$   & $x$                           & $\checkmark$$\checkmark$                  & $\checkmark$                 &                   & $x$                              & $\checkmark$$\checkmark$         & $\checkmark$                     & $x$                           & $\checkmark$$\checkmark$                  & $\checkmark$                  &                   & $x$                              & $\checkmark$$\checkmark$         & $\checkmark$                     & $x$                           & $\checkmark$$\checkmark$                  & $\checkmark$                  &                   & $x$                                   & $\checkmark$             & $\checkmark$$\checkmark$            & $x$                           & $\checkmark$$\checkmark$                  & $\checkmark$                  \\
					& 52                                                    & $x$                                             & $\checkmark$$\checkmark$           & $\checkmark$   & $x$                           & $\checkmark$$\checkmark$                  & $\checkmark$                 &                   & $x$                              & $\checkmark$$\checkmark$         & $\checkmark$                     & $x$                           & $\checkmark$$\checkmark$                  & $\checkmark$                  &                   & $x$                              & $\checkmark$                     & $\checkmark$$\checkmark$         & $x$                           & $\checkmark$$\checkmark$                  & $\checkmark$                  &                   & $\checkmark$                          & $x$                      & $\checkmark$$\checkmark$            & $x$                           & $\checkmark$$\checkmark$                  & $\checkmark$                  \\
					& 252                                                   & $x$                                             & $\checkmark$$\checkmark$           & $\checkmark$   & $x$                           & $\checkmark$$\checkmark$                  & $\checkmark$                 &                   & $x$                              & $\checkmark$                     & $\checkmark$$\checkmark$         & $x$                           & $\checkmark$$\checkmark$                  & $\checkmark$                  &                   & $\checkmark$                     & $x$                              & $\checkmark$$\checkmark$         & $x$                           & $\checkmark$$\checkmark$                  & $\checkmark$                  &                   & $\checkmark$                          & $x$                      & $\checkmark$$\checkmark$            & $x$                           & $\checkmark$$\checkmark$                  & $\checkmark$                  \\\hline
					\multirow{5}{*}{0.7}  & 2                                                     & $\checkmark$                                    & $\checkmark$$\checkmark$           & $x$            & $\checkmark$                  & $\checkmark$$\checkmark$                  & $x$                          & \multirow{5}{*}{} & $\checkmark$                     & $\checkmark$$\checkmark$         & $x$                              & $\checkmark$                  & $\checkmark$$\checkmark$                  & $x$                           & \multirow{5}{*}{} & $\checkmark$                     & $\checkmark$$\checkmark$         & $x$                              & $x$                           & $\checkmark$$\checkmark$                  & $\checkmark$                  & \multirow{5}{*}{} & $x$                                   & $\checkmark$             & $\checkmark$$\checkmark$            & $x$                           & $\checkmark$$\checkmark$                  & $\checkmark$                  \\
					& 4                                                     & $\checkmark$                                    & $\checkmark$$\checkmark$           & $x$            & $\checkmark$                  & $\checkmark$$\checkmark$                  & $x$                          &                   & $\checkmark$                     & $\checkmark$$\checkmark$         & $x$                              & $\checkmark$                  & $\checkmark$$\checkmark$                  & $x$                           &                   & $\checkmark$                     & $\checkmark$$\checkmark$         & $x$                              & $x$                           & $\checkmark$$\checkmark$                  & $\checkmark$                  &                   & $x$                                   & $\checkmark$             & $\checkmark$$\checkmark$            & $x$                           & $\checkmark$$\checkmark$                  & $\checkmark$                  \\
					& 12                                                    & $\checkmark$                                    & $\checkmark$$\checkmark$           & $x$            & $\checkmark$                  & $\checkmark$$\checkmark$                  & $x$                          &                   & $\checkmark$                     & $\checkmark$$\checkmark$         & $x$                              & $\checkmark$                  & $\checkmark$$\checkmark$                  & $x$                           &                   & $\checkmark$                     & $\checkmark$$\checkmark$         & $x$                              & $x$                           & $\checkmark$$\checkmark$                  & $\checkmark$                  &                   & $x$                                   & $\checkmark$             & $\checkmark$$\checkmark$            & $x$                           & $\checkmark$$\checkmark$                  & $\checkmark$                  \\
					& 52                                                    & $\checkmark$                                    & $\checkmark$$\checkmark$           & $x$            & $\checkmark$                  & $\checkmark$$\checkmark$                  & $x$                          &                   & $\checkmark$                     & $\checkmark$$\checkmark$         & $x$                              & $\checkmark$                  & $\checkmark$$\checkmark$                  & $x$                           &                   & $x$                              & $\checkmark$                     & $\checkmark$$\checkmark$         & $x$                           & $\checkmark$$\checkmark$                  & $\checkmark$                  &                   & $\checkmark$                          & $x$                      & $\checkmark$$\checkmark$            & $x$                           & $\checkmark$$\checkmark$                  & $\checkmark$                  \\
					& 252                                                   & $\checkmark$                                    & $\checkmark$$\checkmark$           & $x$            & $\checkmark$                  & $\checkmark$$\checkmark$                  & $x$                          &                   & $x$                              & $\checkmark$                     & $\checkmark$$\checkmark$         & $\checkmark$                  & $\checkmark$$\checkmark$                  & $x$                           &                   & $\checkmark$                     & $x$                              & $\checkmark$$\checkmark$         & $x$                           & $\checkmark$$\checkmark$                  & $\checkmark$                  &                   & $\checkmark$                          & $x$                      & $\checkmark$$\checkmark$            & $x$                           & $\checkmark$$\checkmark$                  & $\checkmark$                  \\\hline
					\multirow{5}{*}{0.8}  & 2                                                     & $\checkmark$                                    & $\checkmark$$\checkmark$           & $x$            & $\checkmark$                  & $\checkmark$$\checkmark$                  & $x$                          & \multirow{5}{*}{} & $\checkmark$                     & $\checkmark$$\checkmark$         & $x$                              & $\checkmark$                  & $\checkmark$$\checkmark$                  & $x$                           & \multirow{5}{*}{} & $\checkmark$                     & $\checkmark$$\checkmark$         & $x$                              & $x$                           & $\checkmark$$\checkmark$                  & $\checkmark$                  & \multirow{5}{*}{} & $x$                                   & $\checkmark$             & $\checkmark$$\checkmark$            & $x$                           & $\checkmark$$\checkmark$                  & $\checkmark$                  \\
					& 4                                                     & $\checkmark$                                    & $\checkmark$$\checkmark$           & $x$            & $\checkmark$                  & $\checkmark$$\checkmark$                  & $x$                          &                   & $\checkmark$                     & $\checkmark$$\checkmark$         & $x$                              & $\checkmark$                  & $\checkmark$$\checkmark$                  & $x$                           &                   & $\checkmark$                     & $\checkmark$$\checkmark$         & $x$                              & $x$                           & $\checkmark$$\checkmark$                  & $\checkmark$                  &                   & $x$                                   & $\checkmark$             & $\checkmark$$\checkmark$            & $x$                           & $\checkmark$$\checkmark$                  & $\checkmark$                  \\
					& 12                                                    & $\checkmark$                                    & $\checkmark$$\checkmark$           & $x$            & $\checkmark$                  & $\checkmark$$\checkmark$                  & $x$                          &                   & $\checkmark$                     & $\checkmark$$\checkmark$         & $x$                              & $\checkmark$                  & $\checkmark$$\checkmark$                  & $x$                           &                   & $\checkmark$                     & $\checkmark$$\checkmark$         & $x$                              & $x$                           & $\checkmark$$\checkmark$                  & $\checkmark$                  &                   & $\checkmark$                          & $x$                      & $\checkmark$$\checkmark$            & $x$                           & $\checkmark$$\checkmark$                  & $\checkmark$                  \\
					& 52                                                    & $\checkmark$                                    & $\checkmark$$\checkmark$           & $x$            & $\checkmark$                  & $\checkmark$$\checkmark$                  & $x$                          &                   & $\checkmark$                     & $\checkmark$$\checkmark$         & $x$                              & $\checkmark$                  & $\checkmark$$\checkmark$                  & $x$                           &                   & $\checkmark$                     & $x$                              & $\checkmark$$\checkmark$         & $x$                           & $\checkmark$$\checkmark$                  & $\checkmark$                  &                   & $\checkmark$                          & $x$                      & $\checkmark$$\checkmark$            & $x$                           & $\checkmark$$\checkmark$                  & $\checkmark$                  \\
					& 252                                                   & $\checkmark$                                    & $\checkmark$$\checkmark$           & $x$            & $\checkmark$                  & $\checkmark$$\checkmark$                  & $x$                          &                   & $x$        & $\checkmark$                              & $\checkmark$$\checkmark$         & $\checkmark$                  & $\checkmark$$\checkmark$                  & $x$                           &                   & $\checkmark$                     & $x$                              & $\checkmark$$\checkmark$         & $x$                           & $\checkmark$$\checkmark$                  & $\checkmark$                  &                   & $\checkmark$                          & $x$                      & $\checkmark$$\checkmark$            & $x$                           & $\checkmark$$\checkmark$                  & $\checkmark$                  \\\hline
					\multirow{5}{*}{0.99} & 2                                                     & \multicolumn{1}{l}{$\checkmark$$\checkmark$}    & \multicolumn{1}{l}{$\checkmark$}   & $x$            & $\checkmark$                  & $\checkmark$$\checkmark$                  & $x$                          & \multirow{5}{*}{} & $\checkmark$$\checkmark$         & $\checkmark$                     & $x$                              & $\checkmark$                  & $\checkmark$$\checkmark$                  & $x$                           & \multirow{5}{*}{} & $\checkmark$$\checkmark$         & $\checkmark$                     & $x$                              & $\checkmark$                  & $\checkmark$$\checkmark$                  & $x$                           & \multirow{5}{*}{} & $\checkmark$$\checkmark$              & $x$                      & $\checkmark$                        & $\checkmark$                  & $\checkmark$$\checkmark$                  & $x$                           \\
					& 4                                                     & \multicolumn{1}{l}{$\checkmark$$\checkmark$}    & \multicolumn{1}{l}{$\checkmark$}   & $x$            & $\checkmark$                  & $\checkmark$$\checkmark$                  & $x$                          &                   & $\checkmark$$\checkmark$         & $\checkmark$                     & $x$                              & $\checkmark$                  & $\checkmark$$\checkmark$                  & $x$                           &                   & $\checkmark$$\checkmark$         & $\checkmark$                     & $x$                              & $\checkmark$                  & $\checkmark$$\checkmark$                  & $x$                           &                   & $\checkmark$$\checkmark$              & $x$                      & $\checkmark$                        & $\checkmark$                  & $\checkmark$$\checkmark$                  & $x$                           \\
					& 12                                                    & \multicolumn{1}{l}{$\checkmark$$\checkmark$}    & \multicolumn{1}{l}{$\checkmark$}   & $x$            & $\checkmark$                  & $\checkmark$$\checkmark$                  & $x$                          &                   & $\checkmark$$\checkmark$         & $\checkmark$                     & $x$                              & $\checkmark$                  & $\checkmark$$\checkmark$                  & $x$                           &                   & $\checkmark$$\checkmark$         & $x$                              & $\checkmark$                     & $\checkmark$                  & $\checkmark$$\checkmark$                  & $x$                           &                   & $\checkmark$$\checkmark$              & $x$                      & $\checkmark$                        & $\checkmark$                  & $\checkmark$$\checkmark$                  & $x$                           \\
					& 52                                                    & \multicolumn{1}{l}{$\checkmark$$\checkmark$}    & \multicolumn{1}{l}{$\checkmark$}   & $x$            & $\checkmark$                  & $\checkmark$$\checkmark$                  & $x$                          &                   & $\checkmark$$\checkmark$         & $\checkmark$                     & $x$                              & $\checkmark$                  & $\checkmark$$\checkmark$                  & $x$                           &                   & $\checkmark$$\checkmark$         & $x$                              & $\checkmark$                     & $\checkmark$                  & $\checkmark$$\checkmark$                  & $x$                           &                   & $\checkmark$$\checkmark$              & $x$                      & $\checkmark$                        & $\checkmark$                  & $\checkmark$$\checkmark$                  & $x$                           \\
					& 252                                                   & \multicolumn{1}{l}{$\checkmark$$\checkmark$}    & \multicolumn{1}{l}{$\checkmark$}   & $x$            & $\checkmark$                  & $\checkmark$$\checkmark$                  & $x$                          &                   & $\checkmark$$\checkmark$         & $x$                              & $\checkmark$                     & $\checkmark$                  & $\checkmark$$\checkmark$                  & $x$                           &                   & $\checkmark$$\checkmark$         & $x$                              & $\checkmark$                     & $\checkmark$                  & $\checkmark$$\checkmark$                  & $x$                           &                   & $\checkmark$$\checkmark$              & $x$                      & $\checkmark$                        & $\checkmark$                  & $\checkmark$$\checkmark$                  & $x$                          
					\\       
					\hline
					
				\end{tabular}
		}}
		\label{tab4}
	\end{table}

	When $\lambda=0.5$ in the presence of transaction costs, the best decision is to trade the wrong asset when $\rho=0.99$ and for all $\rho \leq 0.8,$ the best decision is to trade the right asset, except at 252 time steps. Trading the wrong asset was still the best decision at $\rho=0.99$ even when $\lambda$ decreases to 0.3 and $0.2.$

	\subsubsection{Transaction costs of 1\% on the wrong asset and 2\% on the right asset} 
	We consider a case in which the transaction costs on the right asset are 1\% greater than those on the wrong asset for different $\lambda$ values with the same parameters used in the previous simulation, but with transaction costs on the wrong asset as 1\% and those on the right asset as 2\%. 
	In the absence of transaction costs, the best decision was to trade the right asset for all $\rho$ values at all time-steps. With the inclusion of transaction costs, the best decision to opt for when $\rho=0.99$ is to trade the wrong asset for all $\lambda$ values. When $\lambda=2,$ trading the right asset is the best decision for all $\rho \leq 0.8.$  When $\lambda$ decreases to $0.5,$ the best decision to undertake is to trade the right asset for all $\rho \leq 0.8$ at $2,4$ and 12 time steps and for the rest of the time steps, not hedging at all wins.
	However, when $\lambda=0.3$ and $0.2,$ not hedging  is the best decision for all $\rho \leq 0.8.$

	\subsubsection{ Transaction Costs of 10\% on both the assets}

	A 10\% transaction cost  is  high and can be used in simulations of hedging with transaction costs to help  draw conclusions. Having a 10\% or more transaction costs  indicates that hedging should not be conducted because in the presence of (high) transaction costs, the more one hedge, the more money is lost.
	With these high transaction costs, all Leland's numbers  in the simulations conducted were greater than one because the transaction costs were quite high. This indicates that this  modified delta hedging strategy will not be beneficial in the  presence of  high transaction costs because they reduce its effectiveness.

	In the absence of these high transaction costs, the best decision is to always trade the right asset for all $\rho$ values at all rebalancing intervals; for lower $\rho$ values, the worst decision is to trade the wrong asset. With the inclusion of high and equal transaction costs and $\lambda=2,$ trading the right asset is the best decision for all $\rho$ when time steps are $2,4$ and $12,$ and  the best decision when time steps are 52 and 252 is not to  hedge. For all $\rho \leq 0.8,$ the worst decision is to trade the wrong assets. When $\lambda$ decreases to $0.5,0.3$ and $0.2,$ the best decision is not to trade any of the assets for all $\rho$ values and at all the  rebalancing intervals. A summary of the conclusions for $\rho =0.2$ and 0.99  is presented in Table \ref{tab5}.

	\begin{table}[h]\centering
		\caption{Some of the conclusions from calculated $RAV$ values when $S0_1 = S0_2 =K = 50,~~r = 0.02,~~\sigma_1 = 0.2,~~\sigma_2 = 0.2,~~
			\mu_1 = 0.05,~~\mu_2 = 0.05,$ $\alpha=0,~k_{\text{wrong}}=0.1,~k_{\text{right}}=0.1$}
		{	
			\resizebox{!}{1.49cm}{%
				
				\begin{tabular}{ccccccccccccccccccccccccccccc}
					
					\hline
					
					\multicolumn{2}{c}{\multirow{3}{*}{}}                                         & \multicolumn{6}{c}{ \text{$\lambda=2$}}                                                                               & \multirow{4}{*}{} & \multicolumn{6}{c}{\text{$\lambda=0.5$}}                                                                               & \multirow{4}{*}{} & \multicolumn{6}{c}{\text{$\lambda=0.3$}}                                                                               & \multirow{4}{*}{} & \multicolumn{6}{c}{\text{$\lambda=0.2$}}                                                                               \\ \cline{3-8} \cline{10-15} \cline{17-22} \cline{24-29} 
					\multicolumn{2}{c}{}                                                          & \multicolumn{3}{c}{\multirow{2}{*}{\begin{tabular}[c]{@{}c@{}}\text{With}\\ \text{Transaction Costs}\end{tabular}}} & \multicolumn{3}{c}{\multirow{2}{*}{\begin{tabular}[c]{@{}c@{}}\text{Without}\\ \text{Transaction Costs}\end{tabular}}} &                   & \multicolumn{3}{c}{\multirow{2}{*}{\begin{tabular}[c]{@{}c@{}}\text{With} \\ \text{Transaction Costs}\end{tabular}}} & \multicolumn{3}{c}{\multirow{2}{*}{\begin{tabular}[c]{@{}c@{}}\text{Without} \\ \text{Transaction Costs}\end{tabular}}} &                   & \multicolumn{3}{c}{\multirow{2}{*}{\begin{tabular}[c]{@{}c@{}}\text{With} \\ \text{Transaction Costs}\end{tabular}}} & \multicolumn{3}{c}{\multirow{2}{*}{\begin{tabular}[c]{@{}c@{}}\text{Without} \\ \text{Transaction Costs}\end{tabular}}} &                   & \multicolumn{3}{c}{\multirow{2}{*}{\begin{tabular}[c]{@{}c@{}}With \\ \text{Transaction} \text{Costs}\end{tabular}}} & \multicolumn{3}{c}{\multirow{2}{*}{\begin{tabular}[c]{@{}c@{}}\text{Without} \\ \text{Transaction Costs}\end{tabular}}} \\
					\multicolumn{2}{c}{}                                                          & \multicolumn{3}{c}{}                                                                                  & \multicolumn{3}{c}{}                                                                                     &                   & \multicolumn{3}{c}{}                                                                                   & \multicolumn{3}{c}{}                                                                                      &                   & \multicolumn{3}{c}{}                                                                                   & \multicolumn{3}{c}{}                                                                                      &                   & \multicolumn{3}{c}{}                                                                                   & \multicolumn{3}{c}{}                                                                                      \\ \cline{1-8} \cline{10-15} \cline{17-22} \cline{24-29} 
					\text{$\rho$ }               & \begin{tabular}[c]{@{}c@{}}\text{num\_}\\ \text{steps}\end{tabular} & \text{WA}                            & \text{RA}                                      & \text{NA}                          & \text{WA}                            & \text{RA}                                        & \text{NA}                           &                   & \text{WA}                           & \text{RA}                                       & \text{NA}                           & \text{WA}                            & \text{RA}                                        & \text{NA}                            &                   & \text{WA}                                   & \text{RA}                                   & \text{NA}                       & \text{WA}                            & \text{RA}                                        & \text{NA}                            &                   & \text{WA}                               & \text{RA}                               & \text{NA}                               & \text{WA}                            & \text{RA}                                        & \text{NA}                          \\\hline			\multirow{5}{*}{0.2}  & 2                                                     & $x$                & $\checkmark$$\checkmark$       & $\checkmark$                   & $x$                      & $\checkmark$$\checkmark$             & $\checkmark$           &                   & $x$              & $\checkmark$              & $\checkmark$$\checkmark$              & $x$                      & $\checkmark$$\checkmark$            & $\checkmark$           &                   & $x$              & $\checkmark$              & $\checkmark$$\checkmark$              & $x$                      & $\checkmark$$\checkmark$            & $\checkmark$            &                   & $x$              & $\checkmark$              & $\checkmark$$\checkmark$              & $x$                      & $\checkmark$$\checkmark$            & $\checkmark$            \\
					& 4                                                     & $x$                & $\checkmark$$\checkmark$       & $\checkmark$                   & $x$                      & $\checkmark$$\checkmark$             & $\checkmark$           &                   & $x$              & $\checkmark$              & $\checkmark$$\checkmark$              & $x$                      & $\checkmark$$\checkmark$            & $\checkmark$           &                   & $x$              & $\checkmark$              & $\checkmark$$\checkmark$              & $x$                      & $\checkmark$$\checkmark$            & $\checkmark$            &                   & $x$              & $\checkmark$              & $\checkmark$$\checkmark$              & $x$                      & $\checkmark$$\checkmark$            & $\checkmark$            \\
					& 12                                                    & $x$                & $\checkmark$$\checkmark$       & $\checkmark$                   & $x$                      & $\checkmark$$\checkmark$             & $\checkmark$           &                   & $x$              & $\checkmark$              & $\checkmark$$\checkmark$              & $x$                      & $\checkmark$$\checkmark$            & $\checkmark$           &                   & $x$              & $\checkmark$              & $\checkmark$$\checkmark$              & $x$                      & $\checkmark$$\checkmark$            & $\checkmark$            &                   & $x$              & $\checkmark$              & $\checkmark$$\checkmark$              & $x$                      & $\checkmark$$\checkmark$            & $\checkmark$            \\
					& 52                                                    & $x$                & $\checkmark$                   & $\checkmark$$\checkmark$       & $x$                      & $\checkmark$$\checkmark$             & $\checkmark$           &                   & $x$              & $\checkmark$              & $\checkmark$$\checkmark$              & $x$                      & $\checkmark$$\checkmark$            & $\checkmark$           &                   & $x$              & $\checkmark$              & $\checkmark$$\checkmark$              & $x$                      & $\checkmark$$\checkmark$            & $\checkmark$            &                   & $x$              & $\checkmark$              & $\checkmark$$\checkmark$              & $x$                      & $\checkmark$$\checkmark$            & $\checkmark$            \\
					& 252                                                   & $x$                & $\checkmark$                   & $\checkmark$$\checkmark$       & $x$                      & $\checkmark$$\checkmark$             & $\checkmark$           &                   & $x$              & $\checkmark$              & $\checkmark$$\checkmark$              & $x$                      & $\checkmark$$\checkmark$            & $\checkmark$           &                   & $x$              & $\checkmark$              & $\checkmark$$\checkmark$              & $x$                      & $\checkmark$$\checkmark$            & $\checkmark$            &                   & $x$              & $\checkmark$              & $\checkmark$$\checkmark$              & $x$                      & $\checkmark$$\checkmark$            & $\checkmark$      \\\hline
					\multirow{5}{*}{0.99} & 2                                                     & $\checkmark$       & $\checkmark$$\checkmark$       & $x$                            & $\checkmark$             & $\checkmark$$\checkmark$             & $x$                    & \multirow{5}{*}{} & $x$              & $\checkmark$              & $\checkmark$$\checkmark$              & $\checkmark$             & $\checkmark$$\checkmark$            & $x$                    & \multirow{5}{*}{} & $x$              & $\checkmark$              & $\checkmark$$\checkmark$              & $\checkmark$             & $\checkmark$$\checkmark$            & $x$                     & \multirow{5}{*}{} & $x$              & $\checkmark$              & $\checkmark$$\checkmark$              & $\checkmark$             & $\checkmark$$\checkmark$            & $x$                     \\
					& 4                                                     & $\checkmark$       & $\checkmark$$\checkmark$       & $x$                            & $\checkmark$             & $\checkmark$$\checkmark$             & $x$                    &                   & $x$              & $\checkmark$              & $\checkmark$$\checkmark$              & $\checkmark$             & $\checkmark$$\checkmark$            & $x$                    &                   & $x$              & $\checkmark$              & $\checkmark$$\checkmark$              & $\checkmark$             & $\checkmark$$\checkmark$            & $x$                     &                   & $x$              & $\checkmark$              & $\checkmark$$\checkmark$              & $\checkmark$             & $\checkmark$$\checkmark$            & $x$                     \\
					& 12                                                    & $\checkmark$       & $\checkmark$$\checkmark$       & $x$                            & $\checkmark$             & $\checkmark$$\checkmark$             & $x$                    &                   & $x$              & $\checkmark$              & $\checkmark$$\checkmark$              & $\checkmark$             & $\checkmark$$\checkmark$            & $x$                    &                   & $x$              & $\checkmark$              & $\checkmark$$\checkmark$              & $\checkmark$             & $\checkmark$$\checkmark$            & $x$                     &                   & $x$              & $\checkmark$              & $\checkmark$$\checkmark$              & $\checkmark$             & $\checkmark$$\checkmark$            & $x$                     \\
					& 52                                                    & $x$                & $\checkmark$                   & $\checkmark$$\checkmark$       & $\checkmark$             & $\checkmark$$\checkmark$             & $x$                    &                   & $x$              & $\checkmark$              & $\checkmark$$\checkmark$              & $\checkmark$             & $\checkmark$$\checkmark$            & $x$                    &                   & $x$              & $\checkmark$              & $\checkmark$$\checkmark$              & $\checkmark$             & $\checkmark$$\checkmark$            & $x$                     &                   & $x$              & $\checkmark$              & $\checkmark$$\checkmark$              & $\checkmark$             & $\checkmark$$\checkmark$            & $x$                     \\
					& 252                                                   & $x$                & $\checkmark$                   & $\checkmark$$\checkmark$       & $\checkmark$             & $\checkmark$$\checkmark$             & $x$                    &                   & $x$              & $\checkmark$              & $\checkmark$$\checkmark$              & $\checkmark$             & $\checkmark$$\checkmark$            & $x$                    &                   & $x$              & $\checkmark$              & $\checkmark$$\checkmark$              & $\checkmark$             & $\checkmark$$\checkmark$            & $x$                     &                   & $x$              & $\checkmark$              & $\checkmark$$\checkmark$              & $\checkmark$             & $\checkmark$$\checkmark$            & $x$                    
					\\       
					\hline
					
				\end{tabular}
		}}	\label{tab5}
	\end{table}
	\subsubsection{High and not equal Transaction Costs on both the assets}

	Lastly, we carried out a simulation with the same parameters used in the previous simulations, but with transaction costs on the wrong asset at 10\% and those on the right asset at 20\%.
	In this case, the expected value of no hedging is higher than that of hedging any of the assets owing to the extreme transaction costs on the assets (0.2 on the right asset and 0.1 on the wrong asset). 
	%
	In the absence of transaction costs, the best decision is to trade the right asset, and in the presence of these high transaction costs, the best decision to undertake is not to hedge for all $\lambda$ values and all $\rho$ values. The second-best option is to trade the wrong asset, while the worst decision is to trade the right asset with very high transaction costs.
	
	In summary, the size of $\lambda$ as well as the size of transaction costs on both  assets are key in deciding which asset to trade in the portfolio. For very small transaction costs or no costs on the wrong asset and very small costs on the right asset, $RAV$ concludes that trading the right asset is the best option in the simulations that we conducted. Trading the wrong asset can be opted for when $\rho$ is as high as 0.99 for reasonable transaction costs on the right asset. 
	An increase in the transaction costs on the right asset, and those on the wrong one being smaller, makes the best decision to alternate between never to trade and trading the right asset depending on the size of $\lambda$.
	However, for very high transaction costs, $RAV $ conclusions show that not hedging is the best decision. 
	When transaction  costs on stocks are reasonably small, there is a possibility of trading the wrong asset provided its transaction costs are smaller than those on the right asset for a $\rho$ value as high as $0.99.$

	\subsection{Managerial insights from  trading just one asset in Equation \eqref{eqn1}}\label{subsec:3.3}

	An investor who is considerate of the risk and returns from hedging  chooses to draw conclusions based on the $RAV$ and not just the variance of the simulated values. An investor with two assets, both with reasonable transaction costs,  whereby the one he or she is interested in is so expensive and yields lower returns when traded, has the possibility of trading the wrong asset with lower transaction costs  to obtain higher returns provided there is a very strong positive correlation between the assets. A portfolio of these two assets can be created, provided that they are correlated.
	
	Trading only one asset that is  perfectly or not  correlated with another asset but cheaper to trade enables a financial manager to obtain higher returns than trading an expensive asset that can yield higher risks and lower returns. For this hedging strategy to be effective, an investor should not only consider transaction cost size but also the correlation between the assets in the portfolio and the market price of risk, $\lambda$.
	Our hedging strategy enables the trading of the wrong correlated asset at expiry when the right one becomes too expensive and still makes profits instead of trading the right one and losing money.
	Table \ref{tab6} summarizes our numerical results and indicates when it is best to trade the wrong, right, and no assets.
	For very high transaction costs on both assets $(10\%),$ trading the wrong asset is never the best decision; however strong the correlation between the two assets is.
	
	\begin{table}[h]\centering
		\caption{A summary of the numerical results in presence of transaction costs indicating when it is the best decision to trade the wrong, right asset  and not to trade at all.  
		}
		{ \resizebox{!}{4.5cm}{
				\begin{tabular}{p{1.8cm}p{1.3cm}cp{3.9cm}p{3.3cm}}
					\hline
					\begin{tabular}[c]{@{}c@{}}\text{Transaction}\\ \text{costs}\\ \text{used}\end{tabular}                                                                               & \begin{tabular}[c]{@{}c@{}}\text{Market} \\ \text{price of risk,}\\ \text{$\lambda$}\end{tabular} & \begin{tabular}[c]{@{}c@{}}\text{When is it}\\ \text{best to trade the} \\ \text{wrong asset} \end{tabular} & \begin{tabular}[c]{@{}c@{}}\text{When is it}\\\text{ best to trade the}\\ \text{right asset} \end{tabular} & \begin{tabular}[c]{@{}c@{}}\text{When is it}\\\text{ best to trade} \\ \text{none of the assets}\end{tabular} \\ \hline
					\multirow{4}{*}{\begin{tabular}[c]{@{}c@{}} \\ $k_{\text{wrong}}=0, $\\  $k_{\text{right}}=0.001$\end{tabular}}   & $\lambda=2$                                                                  & never                                                                               & all the time                                                                       & never                                                                                  \\\cline{2-5}
					& $\lambda=0.5$                                                                & when $\rho=0.99$ at 252 steps                                                       & all the time except  when $\rho=0.99$ at 252 steps                                 & never                                                                                  \\\cline{2-5}
					& $\lambda=0.3$                                                                & when $\rho=0.99$ at 252 steps                                                       & all the time except when $\rho=0.99$ at 252 steps                                  & never                                                                                  \\\cline{2-5}
					& $\lambda=0.2$                                                                & when $\rho=0.99$ at all steps                                                       & all the time except when $\rho=0.99$                                               & never                                                                                  \\ \hline
					\multirow{4}{*}{\begin{tabular}[c]{@{}c@{}} \\ $k_{\text{wrong}}=0, $\\ $k_{\text{right}}=0.01$\end{tabular}}    & $\lambda=2$                                                                  & when $\rho=0.99$ at all steps                                                       & when $\rho \leq 0.8$                                                               & never                                                                                  \\ \cline{2-5}
					& $\lambda=0.5$                                                                & when $\rho=0.99$ at all steps                                                       & when $\rho \leq 0.8$                                                               & never                                                                                  \\\cline{2-5}
					& $\lambda=0.3$                                                                & when $\rho=0.99$ at all steps                                                       & when $\rho \leq 0.8$                                                               & never                                                                                  \\\cline{2-5}
					& $\lambda=0.2$                                                                & when $\rho=0.99$ at all steps                                                       & when $\rho \leq 0.8$                                                               & never                                                                                  \\ \hline
					\multirow{4}{*}{\begin{tabular}[c]{@{}c@{}} \\ $k_{\text{wrong}}=0.01, $\\ $k_{\text{right}}=0.02$\end{tabular}} & $\lambda=2$                                                                  & when $\rho=0.99$ at all steps                                                       & when $\rho \leq 0.8$                                                               & never                                                                                  \\  \cline{2-5}
					& $\lambda=0.5$                                                                & when $\rho=0.99$ at all steps                                                       & when $\rho \leq 0.8$ at 2,4 and 12 steps                                           & when $\rho \leq 0.8$ at 52 and 252 steps                                               \\\cline{2-5}
					& $\lambda=0.3$                                                                & when $\rho=0.99$ at all steps                                                       & never                                                                              & when $\rho \leq 0.8$                                                                   \\ \cline{2-5}
					& $\lambda=0.2$                                                                & when $\rho=0.99$ at all steps                                                       & never                                                                              & when $\rho \leq 0.8$                                                                   \\ \hline
					\multirow{4}{*}{\begin{tabular}[c]{@{}c@{}} \\ $k_{\text{wrong}}=0.1, $\\ $k_{\text{right}}=0.1$\end{tabular}}   & $\lambda=2$                                                                  & never                                                                               & all $\rho$ at 2,4 and 12 steps                                                     & all $\rho$at 52 and 252 steps                                                          \\\cline{2-5}
					& $\lambda=0.5$                                                                & never                                                                               & when $\rho \leq 0.8$ at 2,4 and 12 steps                                            & all the time                                                                           \\\cline{2-5}
					& $\lambda=0.3$                                                                & never                                                                               & never                                                                              & all the time                                                                           \\\cline{2-5}
					& $\lambda=0.2$                                                                & never                                                                               & never                                                                              & all the time                                                                           \\ \hline
					\multirow{4}{*}{\begin{tabular}[c]{@{}c@{}} \\ $k_{\text{wrong}}=0.1,$\\ $k_{\text{right}}=0.2$ \end{tabular}}   & $\lambda=2$                                                                  & never                                                                               & never                                                                              & all the time                                                                           \\ \cline{2-5}
					& $\lambda=0.5$                                                                & never                                                                               & never                                                                              & all the time                                                                           \\\cline{2-5}
					& $\lambda=0.3$                                                                & never                                                                               & never                                                                              & all the time                                                                           \\ \cline{2-5}
					& $\lambda=0.2$                                                                & never                                                                               & never                                                                              & all the time 		\\       
					\hline
					
				\end{tabular}

			}
		}\label{tab6}
	\end{table}

	\section{Conclusion}\label{sec:4}
	We conducted delta hedging on a portfolio of two assets with $\alpha$ shares in $S_{t_1}$ and $(1-\alpha)$ shares in $S_{t_2}.$ We hedged the portfolio in the presence and absence of transaction costs, and our results show that hedging in the presence of transaction costs leads to a loss of money, which is expected from any hedging strategy. When $\alpha=0,$ the portfolio is on $S_{t_2}$ and it is the right asset in this case, whereas the wrong one is $S_{t_1}.$ When $\alpha=1,$ the right asset is $S_{t_1}$ and the wrong one is $S_{t_2};$ we  had simulations when the right asset was the only one rebalanced and traded at expiry in the whole portfolio, and another case when the wrong asset was  the only one rebalanced and traded at expiry  in the whole portfolio. 
	To determine  when one is able to trade the wrong asset, right asset, or not even trade in our portfolio in the presence or absence of transaction costs, we hedged our portfolio on simulated data using varying parameters of $\lambda, \rho$, transaction costs and rebalancing intervals. We made decisions based on $RAV.$ 
	
	The  size of $\lambda$ and the size of transaction costs on  the assets are crucial in making a decision on which asset to trade in the portfolio for a given $\rho$ and $\lambda$ value. For very small transaction costs or no costs on the wrong asset, and very small costs on the right asset, $RAV$ concludes that trading the right asset is the best option. Trading the wrong asset can be selected when $\rho$ is as high as $0.99.$
	Not hedging is far better than hedging if the transaction costs are extremely high. These very high transaction costs are mainly found in profitable assets such as real estate, condominiums, and  crypto tokens. For acceptable transaction costs on stocks, there is a possibility of trading the wrong asset, provided its transaction costs are smaller than those on the right asset for a very high positive  $\rho.$ 
	While trading only one of the assets, one asset was at times not perfectly correlated to the other but cheaper to trade; this enabled us to smoothly perturb our strategy from trading in the right, but expensive to rebalance assets, to trading only in the wrong, but cheaper to rebalance asset. 
	This study can be extended by using other models and incorporating additional risk and return measures.

	\section*{Disclosure statement}
	No potential conflict of interest was reported by the authors.
	\section*{Data}
	The SageMath codes that were used to generate the data used are available at the github page:  \url{https://github.com/Erina-Nanyonga/Hedging-against-one-underlying-asset}.

\end{document}